\documentclass[10pt, preprint]{aastex}
\input nearby.defs

\newcounter{thefigs}

\newcounter{thetabs}

\begin{document}

\title{ Physical properties and environments of nearby galaxies}

\begin{abstract}
We review the physical properties of nearby, relatively luminous
galaxies, using results from newly available massive data sets
together with more detailed observations. First, we present the global
distribution of properties, including the optical and ultraviolet
luminosity, stellar mass, and atomic gas mass functions. Second, we
describe the shift of the galaxy population from ``late'' galaxy types
in underdense regions to ``early'' galaxy types in overdense regions.
We emphasize that the scaling relations followed by each galaxy type
change very little with environment, with the exception of some minor
but detectable effects. The shift in the population is apparent even
at the densities of small groups and therefore cannot be exclusively
due to physical processes operating in rich clusters. Third, we divide
galaxies into four crude types --- spiral, lenticular, elliptical, and
merging systems --- and describe some of their more detailed
properties. We attempt to put these detailed properties into the
global context provided by large surveys.

\end{abstract}

\author{Michael R. Blanton and John Moustakas}

\affil{Center for Cosmology and Particle Physics, Department of
Physics, \\
New York University, 4 Washington Place, New York, NY 10003 \label{NYU}}

\email{michael.blanton@gmail.com}

\maketitle
\tableofcontents

\section{New windows on the nearby Universe} 

As early as the work of \citet{hubble36a}, astronomers recognized the
existence of distinct galaxy types --- smooth ``early-types''
preferentially found in groups and clusters and complex-looking
``late-types'' preferentially found in less dense regions of the sky.
Despite this long history, we have not yet determined with certainty
the physical mechanisms that differentiate galaxies into classes.  The
past decade of astronomers' effort has yielded both massive new
surveys of the nearby Universe and more detailed observations of
individual objects.  Although many of these observations have yet to be
digested and fully understood, already we have a clearer view of the
detailed physical properties than we did only a decade ago. Focusing
on nearby galaxies comparable in mass to the Milky Way, we review some
of the latest results on the census of the galaxy population.

Recently developed observational tools for understanding galaxy
formation fall into two general types: wide-field surveys and targeted
(but more detailed) observations.  The wide field surveys we focus on
are the Galaxy Evolution Explorer in the ultraviolet (\emph{GALEX};
\citealt{martin05a}), the Sloan Digital Sky Survey in the optical
(SDSS; \citealt{york00a}), the Two-Micron All Sky Survey in the near
infrared \citep[2MASS;][]{skrutskie06a}, as well as 21-cm radio
surveys such as the \HI\ Parkes All Sky Survey (HIPASS;
\citealt{meyer04a}) and ALFALFA on Arecibo (\citealt{giovanelli07a}).
Each of these surveys covers a substantial fraction of the sky with
imaging; redshifts from the SDSS and the 21-cm surveys then provide a
third dimension.  Along with the Two-degree Field Galaxy Redshift
Survey (2dFGRS; \citealt{colless01a}), the Six-degree Field Galaxy
Redshift Survey (6dFGRS; \citealt{jones04a}) and the 2MASS Redshift
Survey (\citealt{crook07a}), these massive surveys supply a detailed
map of the galaxy density field and the framework of large-scale
structure within which galaxies evolve. In addition, they supply a
host of spectroscopic and photometric measurements for each galaxy:
luminosities, sizes, colors, star-formation histories, stellar masses,
velocity dispersions and emission line properties.

The second type of tool consists of targeted but more detailed
programs that are too expensive to conduct on massive scales right
now, but for which even small numbers of galaxies can be
revealing. Such programs include the Spectroscopic Areal Unit for
Research on Optical Nebulae (SAURON; \citealt{bacon01a}), the
\emph{Spitzer} Infrared Nearby Galaxy Survey (SINGS;
\citealt{kennicutt03a}), The \HI\ Nearby Galaxy Survey (THINGS;
\citealt{Walter:2008p2836}), new large Hubble Space Telescope
(\emph{HST}) programs, and large compilations of individual efforts
such as detailed radio observations, long-slit spectroscopy, and deep
optical and near-infrared imaging.

Using the results of these recent efforts, we begin by exploring the
global demographics of galaxies and their dependence on
environment. Then, dividing galaxies into classes (spirals,
lenticulars, ellipticals, and mergers), we review recent results
concerning the scaling relations, star-formation histories, and other
properties of each class.

We cannot hope to be exhaustive, and instead focus on recent results
rather than historical ones. We omit discussion of dwarf systems,
which are rather different than their more massive counterparts, being
generally more gas-rich, disk-dominated, and usually lacking in spiral
structure (a topic ripe for review; meanwhile, see
\citealt{geha06b}). Because of our focus on global properties, we also
deemphasize central black holes (\citealt{kormendy04b}) and active
galactic nuclei (AGN; \citealt{ho08a}), which might have an important
influence on galaxy evolution as a whole (\citealt{kauffmann03b,
  Best:2006p1779, Khalatyan:2008p1776}). For elliptical galaxies, we
refer the reader to multiple other reviews examining their more
detailed properties, including their stellar populations
(\citealt{renzini06a}), their structure and classification
(\citealt{Kormendy:2008p1880}), and their hot gas content
(\citealt{Mathews:2003p1939}). A final warning is that we adopt some
of the language of morphology (elliptical or ``E'', lenticular or
``S0'', and spiral galaxies) without fully addressing the problem of
classification (\citealt{Sandage:2005p1910}).

Throughout, we assume a standard cosmology of $\Omega_m=0.3$ and
$\Omega_\Lambda=0.7$, with $H_0=100 h$ km~s$^{-1}$~Mpc$^{-1}$.  All
magnitudes are on the AB system unless otherwise specified.  

\section{A global view of galaxy properties}
\label{global}

A breakthrough of recent surveys has been the ability to explore many
dimensions of galaxy properties simultaneously and homogeneously, in
order to put galaxy scaling relationships in context with respect to
one another. In this section, we describe these general properties,
their distribution, and their dependence on environment. We
concentrate in this section primarily but not exclusively on SDSS
results, which currently yield the most homogeneous and consistent
measurements for the broadest range of galaxy varieties. In
particular, in the subsections below we will make use of 77,153
galaxies with $z<0.05$ in the SDSS Data Release 6 (DR6;
\citealt{adelman08a}), an update of the
low-redshift sample of \citet{blanton05a}.

\subsection{Optical broad-band measurements }
\label{broadband}

\begin{figure}
\plotone{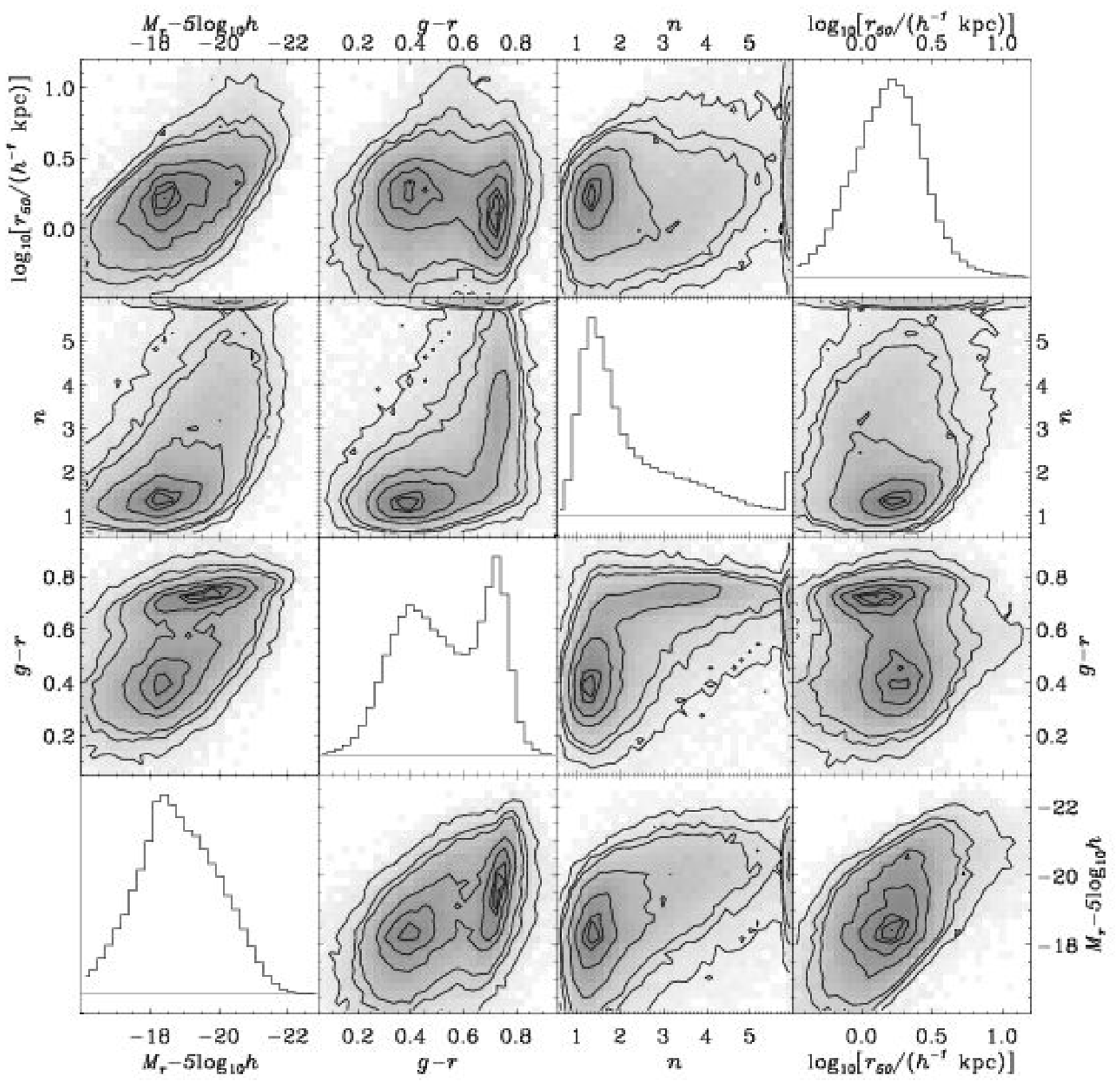}
\caption{\label{manyd} Distribution of broad-band galaxy properties in
the SDSS. The diagonal panels show the distribution of four
properties independently: absolute magnitude $M_r$, $g-r$ color,
\Sersic\ index $n$, and half-light radius $r_{50}$. A bimodal
distribution in $g-r$ is apparent. The off-diagonal panels show the
bivariate distribution of each pair of properties, revealing the
complex relationships among them. The greyscale and contours reflect
the number of galaxies in each bin (darker means larger number).}
\end{figure}

Figure \ref{manyd} shows the simplest measurable properties of
galaxies from the SDSS sample: the absolute magnitude $M_r$, the $g-r$
color, the \citet{sersic68a} index $n$ in the $r$-band, and the
physical half-light radius $r_{50}$ (sometimes called the ``effective
radius''). These properties reveal a variety of correlations, most
known for many years, but now quantified much more precisely.

The absolute magnitude $M_r$ is a critical quantity, correlating well
with stellar mass as well as with dynamical mass (see the discussion
of the Tully-Fisher relation in \S\ref{tully-fisher} and the
fundamental plane in \S\ref{fp}). Clearly, the other properties are a
strong function of overall mass. Although Figure
\ref{manyd} shows the raw distribution for the flux-limited SDSS sample,
in \S\ref{lfandsf} we correct for selection effects and calculate the
luminosity and stellar mass functions.

In many of these plots, particularly those involving $g-r$ color,
there is a bimodal distribution --- galaxies can be divided very
roughly into red and blue sequences (\citealt{strateva01a, blanton03d,
  baldry04a}).  The red and blue classification is not always related
in a simple way to classical morphology --- though of course there is
some relationship (\citealt{roberts94a}).  In particular, galaxies in
the blue sequence are very reliably classifiable as spiral galaxies
with ongoing star-formation (\S\ref{sec:spirals}).  However, the red
sequence contains a mix of types. The lower luminosity end consists of
compact ellipticals (cEs) and dwarf ellipticals (dEs; sometimes known
as spheroidals, Sph). Around $M_r-5\log_{10}h \sim -20$, the red
sequence is a mix of early-type spirals, dust-reddened spirals
(\S\ref{dust}), lenticulars (S0s; \S\ref{sec:lenticulars}), and giant
ellipticals (Es; \S\ref{sec:ellipticals}).  At the highest
luminosities, it consists of cD galaxies (\S\ref{cD}).

For a rough quantification of how these types populate the red
sequence, we use the classifications of our sample galaxies stored in
the NASA Extragalactic Database (NED). In practice, most of these
classifications come from The Third Reference Catalog of Bright
Galaxies (RC3; \citealt{devaucouleurs91a}). For our purposes we select
galaxies within $\Delta(g-r) \sim 0.03$ of the red sequence, with no
detected lines associated with star-formation
(see \S\ref{spectra} and Figure \ref{manyd-spectro}). For any
luminosity $M_r-5\log_{10}h<-17$, only about 40\% of these galaxies
are Es. About 25\%--50\% of them are S0s, with the lowest fractions at
around $M_r - 5\log_{10} h \sim -20$, increasing to both higher and
lower luminosities. We suspect these fractions are in practice
overestimates, since spiral systems are far more commonly
misclassified as E/S0s than vice-versa. It remains clear, however,
that restricting to red sequence galaxies does not suffice to
guarantee an E/S0 sample --- at $M_r - 5\log_{10} h \sim -20$ at least
one-third of the red sequence population is Sa or later.  

Both the red and blue sequences have mean colors that are a function
of absolute magnitude. Blue galaxies have recent star-formation, and
their color is strongly related to the recent star-formation history 
(\S\ref{sfr}) and the dust reddening in the galaxy (\S\ref{dust}).  As
galaxy mass increases, the change in color reflects both the increased
reddening due to dust and the decreased fraction of recent
star-formation (related to an increased importance of red central
bulges; \S\ref{bulge}).
The strong dependence on recent star formation history translates into
the large range of colors on the blue sequence.  As one might expect
given this trend, at high masses the atomic and molecular gas fractions
are low (\S\ref{gas}), whereas the metallicities are high
(\S\ref{yeff}).

Red galaxies have (generally) little recent star-formation, and their
color is weakly related to both mean stellar age and to metallicity,
both of which rise with mass (\S\ref{stellar-pops}).  Because the
dependence of color on these properties is weak relative to their
intrinsic variation, there is a small range of red galaxy colors in
Figure \ref{manyd}. Although their old stellar age is related to their
relatively gas-poor nature, they are not entirely devoid of cold gas
(\S\ref{coldgas}) or perhaps of relatively recent star-formation
(\S\ref{stellar-pops}).

The \Sersic\ index $n$ is a measurement of the overall profile
``shape,'' and is defined by the profile:
\begin{equation}
I(r) \propto \exp\left[-\left(\frac{r}{r_0}\right)^{1/n}\right],
\end{equation}
where $r_0$ is a scale factor that one can relate to the half-light
radius $r_{50}$ given $n$ (\citealt{sersic68a}; see
\citealt{Graham:2005p2523} for the mathematical relationship).  In general,
this model is usually generalized to non-axisymmetric cases by
allowing for an axis-ratio $b/a<1$ (that is, elliptical rather than
circular isophotes).
While $r_{50}$ reflects the physical size of the galaxy, $n$ reflects
what we define as the ``concentration.'' As $n\rightarrow 0$, the
\Sersic\ profile approaches a uniform disk of light with radius
$r_0$. As $n$ increases, the \Sersic\ function simultaneously
concentrates the surface brightness profile towards the center and
pushes flux further out, passing through a Gaussian ($n=0.5$), an
exponential ($n=1$), a de Vaucouleurs profile ($n=4$;
\citealt{devaucouleurs48a}), and even higher for some galaxies. This
index often is used as a proxy for morphology (e.g., \citealt{bell03a,
  shen03a, mandelbaum06a}), though it is only partially related to
classical morphological determinations (\S\ref{morphology}).

For many galaxies a single \Sersic\ index model does not explain the
profile completely. For elliptical galaxies, the central regions are
often not well fit by such a model (\S\ref{core}). For spiral and
lenticular galaxies a better model is an exponential disk plus a
\Sersic\ bulge (\S\ref{bulge}).  However, as \citet{graham01a} shows,
measures of the overall concentration are well-correlated with
bulge-to-disk ratios for real galaxies. 

Consequently, in Figure \ref{manyd} the single component \Sersic\ $n$
does reveal some interesting trends. Blue galaxies generally have low
\Sersic\ indices, while red galaxies span a range of \Sersic\
indices. Overall, both blue galaxies and red galaxies tend to be more
concentrated at higher luminosity. For blue galaxies this reflects an
increased importance of the bulge, and a simultaneous increase in
concentration of the bulge itself (\S\ref{bulge}).  For red galaxies
the trend appears to reflect an overall structural difference between
low luminosity and high luminosity early-type galaxies
(\S\ref{structure}).

Finally, the plots also demonstrate that the optically emitting
regions of galaxies are larger for more massive galaxies.  Luminous
galaxies achieve $r_{50} > 10$ $h^{-1}$ kpc, while the low luminosity
population typically has $r_{50} < 1$ $h^{-1}$ kpc. The full extent of
the galaxies in the optical tends to be at least 2--4 $\times$
$r_{50}$ for most systems depending on the concentration. For disk
galaxies, even this optical radius does not trace the outer gas disk
often visible in \HI\ and in very sensitive ultraviolet (UV)
measurements (\S\ref{xuv}).

\subsection{Optical spectroscopic measurements}
\label{spectra}

\begin{figure}
\plotone{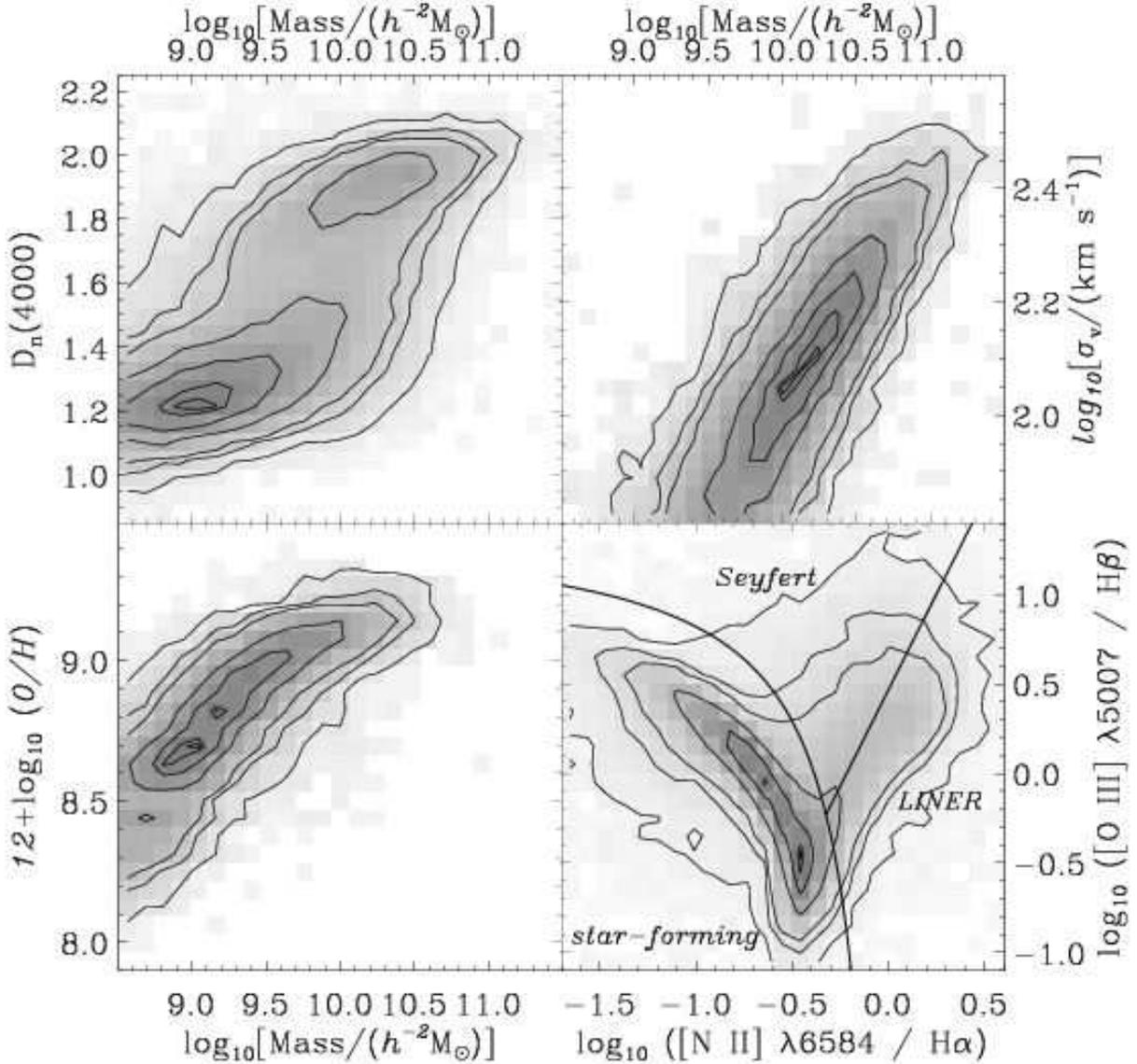}
\caption{\label{manyd-spectro} 
Distribution of spectroscopic galaxy properties in the SDSS (data
courtesy \citealt{brinchmann04a, tremonti04a}). Greyscale and
contours are similar to those in Figure \ref{manyd}. Upper left panel
shows the distribution of stellar mass and $D_n$(4000), a measure of
the stellar population age.  As in the color-magnitude diagram, the
separation between old and young populations is apparent. Upper right
panel shows for red galaxies the distribution of stellar mass and
velocity dispersion, revealing the Faber-Jackson relation. Bottom left
panel shows for star-forming galaxies the relationship between mass
and gas-phase oxygen abundance, the ``mass-metallicity'' relation.
Bottom right panel shows the
\citet{baldwin81a} relationship for emission-line galaxies. Shown are
the divisions (based on \citealt{kauffmann03b}) among star-forming,
Seyfert, and LINER galaxies.}
\end{figure}

More detailed information on stellar populations and the interstellar
medium can be obtained from integrated spectrophotometry 
\citep{kennicutt92a, jansen00a, moustakas06b}.  
In Figure \ref{manyd-spectro}, we show a small sampling of the major
trends in galaxy spectroscopic properties in the optical, using the
same sample as in Figure
\ref{manyd}, with parameters determined by \citet{brinchmann04a} and
\citet{tremonti04a} for our SDSS sample.  The SDSS spectra are taken
through 3~arcsec diameter fibers, a generally small radius within
nearby galaxies, and so aperture effects are not ignorable; however,
the general conclusions we draw here are not affected much by this
consideration.

For the purposes of these plots, we replace luminosity by the stellar
mass. There are numerous methods for calculating stellar mass; see the
compilation of major techniques in \citet{baldry08a}. Here we use the
simple technique of \citet{bell03a}, which assigns a mass-to-light
ratio according to the galaxy broad-band colors. This technique is
robust at least for most galaxies, because in practice increasing
dust, age, and metallicity all both increase the mass-to-light ratio
and redden the colors, in roughly similar proportions. Of course,
galaxies with large, recent bursts of star-formation or extreme
amounts of dust attenuation will not be well-described by this simple
technique \citep{bell01b}.  But the greater unknown in determining
stellar masses is the initial mass function (IMF) of stars assumed,
because the lowest mass stars contribute considerable mass but very
little light.  In fact, as
\citet{baldry08a} show, a number of disparate techniques for
calculating stellar masses agree well at fixed IMF. The
\citet{bell03a} estimate we use employs the non-physical ``diet''
Salpeter IMF, which yields stellar masses slightly higher
($\sim0.05$~dex) than the popular \citet{kroupa01a} and
\citet{chabrier03a} IMFs.

The upper left panel of Figure \ref{manyd-spectro} shows the
distribution of stellar mass and the quantity $D_n$(4000), a measure
of the 4000-\AA\ break, according to the ``narrow'' definition of
\citet{balogh99a} and using the determination of
\citet{brinchmann04a}. As \citet{kauffmann03a} point out, the
4000-\AA\ break traces stellar population age better than broad-band
colors, and it is
less sensitive to dust reddening. Star-forming galaxies tend to have
weak breaks, with $D_n$(4000)$\sim 1.1$--$1.4$, whereas passive stellar
populations tend to have strong breaks, with $D_n$(4000)$\sim
1.8$--$2.1$. Indeed, similar to galaxy color, $D_n$(4000) clearly
separates these two populations; older populations dominate
at high stellar mass and the star-forming systems dominate at low
stellar mass. Based on this diagram, \citet{Kauffmann:2003p2739}
define a critical stellar mass of $1.5\times 10^{10}$ $h^{-2}$
$\mathcal{M}_\odot$ separating these two galaxy populations.

From SDSS spectra, we can also determine galaxy velocity dispersions;
for a sample of galaxies without emission lines, we show the results
in the upper right panel of Figure \ref{manyd-spectro}. Here we see a
clear correlation between velocity dispersion and stellar mass,
related of course to the classic relation of \citet{faber76a}. This
relationship is merely a projection of the fundamental plane
(\S\ref{fp}).  Both the fundamental plane and the Tully-Fisher
relation (\S\ref{tully-fisher}) show that the dynamical masses of
galaxies are correlated with their stellar masses.

The bottom right panel of Figure \ref{manyd-spectro} shows the ``BPT''
diagram (\citealt{baldwin81a}), based on the H$\alpha$ and H$\beta$
recombination lines and the [N~{\sc ii}]~$\lambda$6584 and [O {\sc
    iii}]~$\lambda$5007 collisionally excited forbidden lines. 
The position of an object in the BPT diagram is a function of
metallicity and the 
ionization state of the emitting gas (e.g, \citealt{kennicutt00a}) and
can be roughly divided into three regions (following
\citealt{kauffmann03b}). The lower left region contains emission line
galaxies dominated by star-formation. The upper triangle contains
Seyfert galaxies, which have higher ionizing fluxes and are likely
associated with AGN. The lower right contains Low Ionization Nuclear
Emission-line Regions (LINERs; \citealt{heckman80a, kewley06a}), also
most likely associated with AGN. \citet{ho08a} discuss the
classification and physical properties of nearby AGN in much greater
detail than possible here.

Finally, we isolate the star-forming galaxies (those with emission
lines falling in the star-formation region of the BPT diagram).  The
lower left panel of Figure \ref{manyd-spectro} shows their gas-phase
metallicity (from \citealt{tremonti04a}). We use the standard
quantification of oxygen abundance, $12+\log_{10}{\rm (O/H)}$.
Clearly more massive galaxies are more metal-rich, saturating around
9.2~dex (about 0.5~dex more oxygen-rich than the newly revised solar
abundance; \citealt{melendez08a}).  Closed-box chemical evolution
models do predict an increase in metallicity as the fraction of gas
turned into stars rises, since more generations of star-formation have
occurred.  However, in fact, the observed increase in metallicity is
likely also driven by metal-rich outflows in low-mass galaxies, or
another violation of the closed-box model (\S\ref{yeff}).

\subsection{Luminosity and stellar mass functions}
\label{lfandsf}

A fundamental measurement of the galaxy distribution is the luminosity
function (LF): the number density as a function of luminosity. With
flux-limited samples, such as those shown in Figures \ref{manyd} and
\ref{manyd-spectro}, one needs to apply corrections for the fact that
faint galaxies cannot be observed throughout the survey volume.  A
number of statistical techniques have been developed over the years to
correct for this effect, the simplest of which is the
$1/V_{\mathrm{max}}$ method (\citealt{schmidt68a}).  Using this
method, one counts each galaxy with a weight equal to the inverse of
the volume over which it could have been observed. In the integral
over volume, one must include any weighting for survey completeness as
well as flux or other selection effects, based on the galaxy's
particular properties.  Then, in any bin of any set of properties one
can calculate the number density of galaxies in that bin as $\sum_i
1/V_{\mathrm{max},i}$. For modern samples, the results from this
method agree with others designed to be insensitive to accidental
correlations between large-scale structure and redshift
(e.g. \citealt{efstathiou88a, takeuchi00a}).

\begin{figure}
\plotone{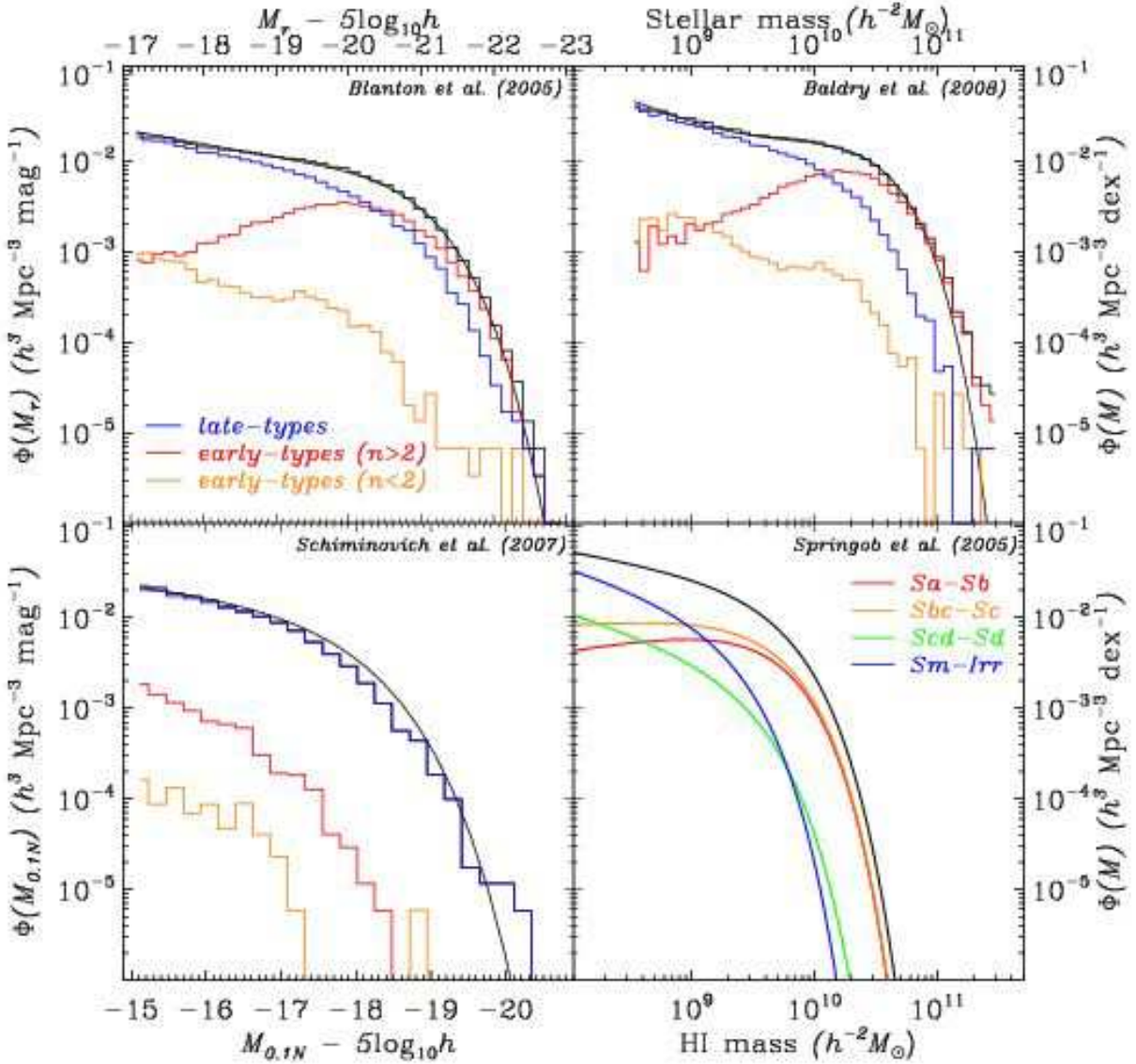}
\caption{\label{lf} Optical and near-UV luminosity function, stellar mass
function, and H~{\sc i} mass function of galaxies.  The upper left
panel shows the local $r$-band luminosity function of all galaxies
(black histogram), as well as the luminosity functions of late-types
(blue), concentrated early-types (red) and diffuse early-types
(orange), as described in \S\ref{lfandsf}. Overplotted as the smooth
curve is the double Schechter function fit of
\citet{blanton04b}. The upper right panel shows the local stellar mass
function for the same galaxies. The bottom left panel shows the
\emph{GALEX} near-UV luminosity function for the various galaxy types 
and the smooth fit to the full luminosity function
from \citet{schiminovich07a}.  Finally, the lower right panel shows
the H~{\sc i} mass function fits from \citet{springob05b} for all
galaxies and as a function of morphological type.}
\end{figure}

The upper left panel of Figure \ref{lf} shows the optical $r$-band LF
from the SDSS DR6 sample, using $V_{\mathrm{max}}$ calculations
described in \citet{blanton04b}. We show both the overall LF and the
LF split among galaxy types, using some very basic criteria.  The
``late'' population consists of any blue galaxies plus any
star-forming red galaxies.  The ``early'' population consists of other
galaxies, split into concentrated early-types, which have $n>2$, and
diffuse early-types, with $n<2$. These classifications are crude, of
course --- this method describes many legitimate Sa galaxies and
basically all S0 galaxies as ``early-type,'' for example.

Nevertheless, this panel begs comparison with Figure 1 of
\citet{binggeli88a}, who show very similar trends in the LF using
visual morphological classifications. As they found, early type
galaxies dominate the bright-end, while late-types dominate the faint
end.  Diffuse early-types 
become increasingly important at low luminosity.

As \citet{binggeli88a} also showed, a \citet{schechter76a} function
fit to the LF is insufficient.  We overplot the double-Schechter
function fit of \citet{blanton04b}, which uses a broken power-law for
the faint end: 
\begin{equation}
\Phi(L) dL = \frac{dL}{L_\ast}
  \exp(-L/L_{\ast}) \left[
\phi_{\ast,1} 
\left( \frac{L}{L_{\ast}} \right)^{\alpha_1} + 
\phi_{\ast,2} 
\left( \frac{L}{L_{\ast}} \right)^{\alpha_2} 
  \right]
\end{equation}
The parameter $L_\ast$ is a fundamental one, indicating the
approximate onset of the exponential cut-off. At the faint-end the
luminosity function is approximately a power law with slope $\alpha_2$
ranging between $-1.35$ and $-1.52$ in the $r$-band depending upon
assumptions about the low surface brightness population
(\S\ref{systematics}).

The upper right panel of Figure \ref{lf} shows the stellar mass
function, using the same stellar mass determinations as used in Figure
\ref{manyd-spectro}. Because the stellar mass-to-light ratios of red
galaxies are higher than blue galaxies, the stellar mass function
accentuates the distinction between the blue and red populations: the
early-types become even more dominant. Overlaid is the double
Schechter fit of \citet{baldry08a}, who present a comprehensive
treatment of recent estimates of the stellar mass function (the
results shown here use the \citealt{chabrier03a} IMF). After
accounting for differences in the adopted IMF, these results
are in rough agreement with determinations of
\citet{cole01a}, \citet{kochanek01a}, and \citet{bell03a} based on 
2MASS $K$-band data.

The lower left panel of Figure \ref{lf} shows the near-UV luminosity
function, which traces recent star-formation, obtained by matching the
SDSS sample to the \emph{GALEX} Release 3 (GR3;
\citealt{martin05a}). We separate the galaxies according to the same
classifications as used above. We $K$-correct here to the near-UV band
shifted blueward by a factor $1+z=1.1$ (\citealt{blanton06b}) for
consistency with \citet{schiminovich07a}. Overlaid as the smooth black
line is their full luminosity function.  Clearly the early-types are
very subdominant --- the UV luminosity of the Universe is completely
dominated by blue, star-forming galaxies.

Finally, in the lower right panel of Figure \ref{lf} we show the \HI\
mass functions determined by the compilation of \citet{springob05b},
which is comparable to the HIPASS result of \citet{zwaan05a}. Since
they did not publish their full functions, we only show the Schechter
function fits (which are accurate descriptions). They limited their
sample to spiral galaxies; most ellipticals have very low atomic gas
content and would not contribute significantly to this plot
(\S\ref{coldgas}), though they often have significant hot ionized gas
(\citealt{Mathews:2003p1939}). They split the sample according to
morphological type (though virtually all of the types would fall into
the ``late-type'' histograms in the other plots). The most notable
effect is that the mass cut-off scale for the \HI\ mass function is
far lower than for the stellar mass function: the most massive
galaxies are dominated by stars, not neutral hydrogen. That situation
changes dramatically at lower masses (\S\ref{gas}).

\subsection{Systematic effects in luminosity functions }
\label{systematics}

For the luminosity and stellar masses in this section, we relied upon
SDSS Petrosian magnitudes (\citealt{petrosian76a, blanton01a,
  strauss02a}), which have the virtue that they are roughly
redshift-independent quantities.  Some researchers advocate the use of
``total'' flux, using extrapolated radial profiles. The difference is
largest for de Vaucouleurs profiles (with SDSS Petrosian fluxes less
by 10\%--20\%, depending on the ratio of the half-light radius to the
seeing).  For very detailed work, such as the build-up of stellar mass
on the red sequence, it is therefore necessary to know the flux
estimator used (e.g., \citealt{brown06a}). In the context of SDSS,
another problem exists: the biggest galaxies on the sky have their
fluxes significantly underestimated owing to oversubtraction of the
background (\citealt{blanton05b, bernardi07a}), yielding a 20\% bias
at $r_{50}\sim 20''$ and rising at larger sizes
(\citealt{lauer07a}). For that reason, the effect on the luminosity
function is minimal, since virtually all of the brightest galaxies in
the sample are far away and thus small in angular size.

All luminosity and stellar mass function estimates are affected at
some level by surface brightness selection effects. \citet{blanton05a}
estimate that for the SDSS these effects become important at the 10\%
level at $M_r -5\log_{10} h \sim -17$ or so. At lower luminosities,
the SDSS spectroscopic survey starts to become incomplete; for
2MASS-based surveys like \citet{cole01a} or \citet{kochanek01a}, the
incompleteness probably sets in at higher luminosities, though no
quantitative estimate exists. A less traditional view is that of
\citet{oneil04a}, who suggest that a substantial reservoir of baryons
could be in massive low surface brightness galaxies, with large H~{\sc
  i} masses ($>10^{10} \mathcal{M}_\odot$) but barely detectable in
optical surveys.  Such galaxies exist, but no quantitative estimate of
their number density has been made.

\subsection{Environmental dependence}
\label{environs}

As has long been established, all of these properties are a strong
function of the local environment --- whether a galaxy is in a cluster
or in a void. With the latest data sets, we now have a much more
detailed understanding of this dependence. To illustrate these
effects, we estimate environment using the number $N_n$ of neighboring
galaxies with $M_r-5\log_{10}h<-18.5$, within a projected distance of
500 $h^{-1}$ kpc and a velocity of $600$ km~s$^{-1}$. 
The luminosity threshold is roughly that of the Large
Magellanic Cloud. Other measures of galaxy environment have been
explored recently: group environment (\citealt{yang07a}), nearest
neighbor distance (\citealt{park07a}), and kernel density smoothing
(\citealt{balogh04a}). The results we describe here seem to hold no
matter which estimate of density one uses.

\begin{figure}
\plotone{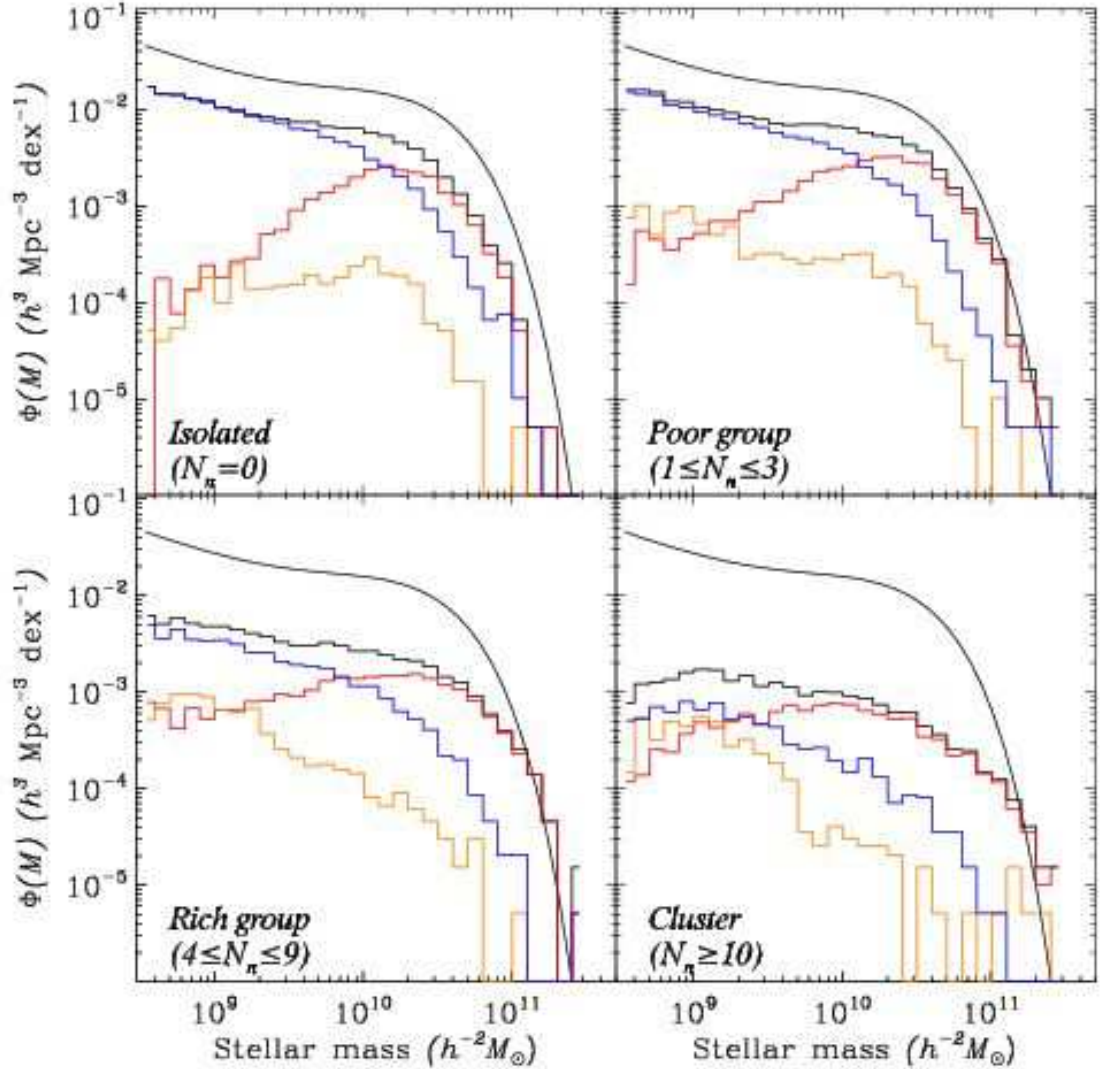}
\caption{\label{lfden} Stellar mass function of
galaxies as a function of morphological type (see Fig.~\ref{lf}
and \S\ref{lfandsf}) and local galaxy density.  Each panel corresponds
to a different environment, ranging from isolated, to poor groups,
rich groups, and clusters, based on the number of neighbors $N_{n}$
with $M_r-5\log_{10}h<-18.5$ within 500~$h^{-1}$~kpc and
600~km~s$^{-1}$.  For reference, this luminosity threshold roughly
corresopnds to the luminosity of the Large Magellanic Cloud.
The double Schechter function fit to all galaxies
from \citet{baldry08a} is shown in all panels as a smooth black
curve.} 
\end{figure}

First, let us consider the variation of the stellar mass function with
environment, as shown in Figure \ref{lfden}. The panels are broken up,
as labeled, into bins of $N_n$ --- the upper right panel is roughly
the environment of galaxies in the Local Group.  The first notable
fact is that most galaxies with $0.01 L_\ast \simless L\simless
L_\ast$ are in relatively underdense environments, at the scale of the
Local Group or less. Generally speaking, the massive galaxies are
relatively more likely to exist in dense regions, reflecting the
variation of the shape of the optical luminosity function from void
regions (\citealt{hoyle03a}), through average regions
(\citealt{blanton05a}), to clusters (\citealt{popesso05a}).

Even in the least dense environments, the ``early-type'' galaxies are
a substantial population --- and indeed dominant at high masses. At
lower masses the population is completely dominated by blue disk
galaxies. As density increases, the characteristic mass of the red
population increases.
Clearly most of the
brightest objects are red and in dense regions.  In addition, the blue
population declines in importance as density increases, becoming
almost outnumbered even at the faint end.

\begin{figure}
\plotone{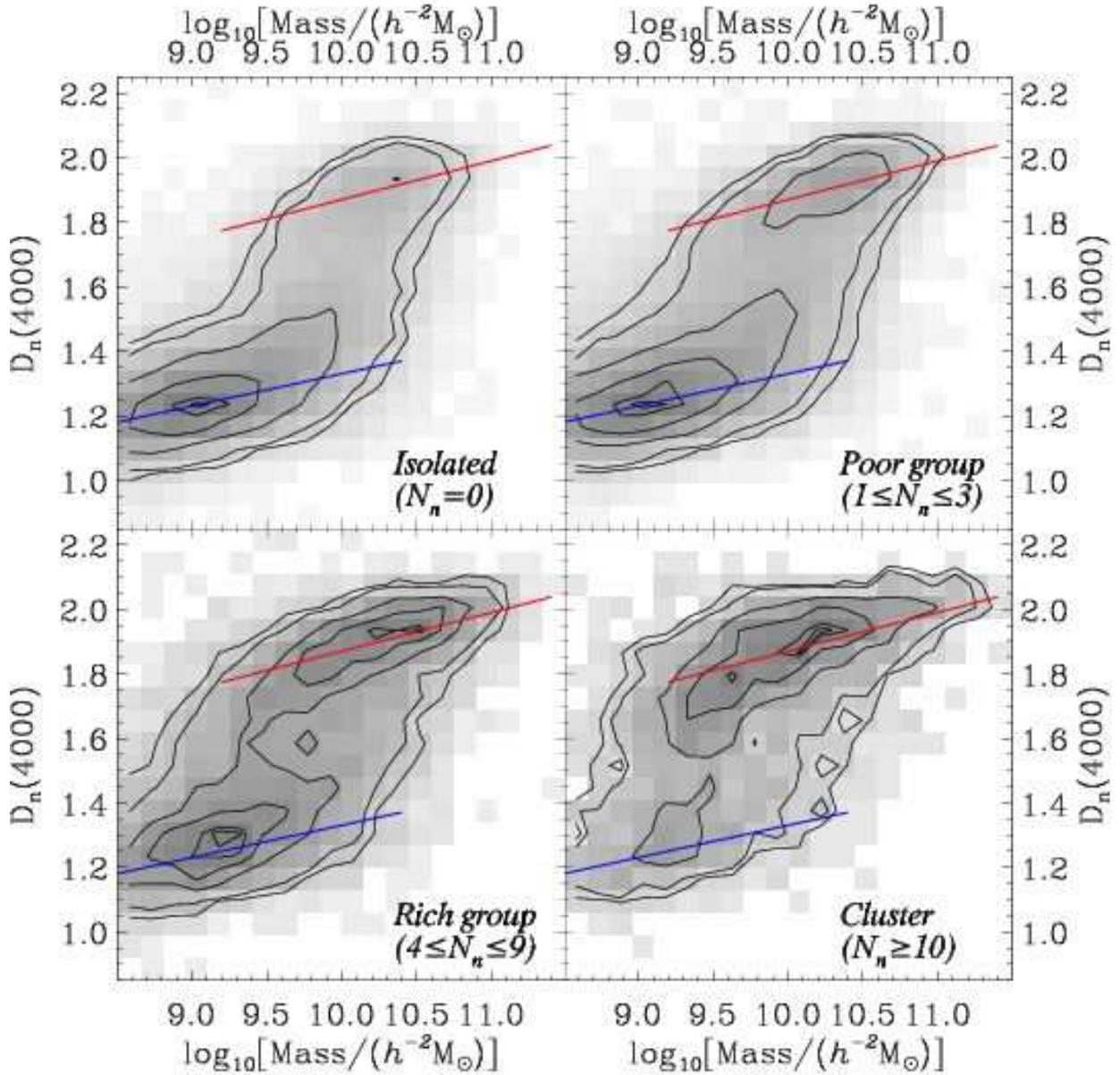}
\caption{\label{d4000_env} Distribution of $D_n$(4000) and stellar
mass as a function of environment as defined in \S\ref{lfden}.  
In each panel we show references for the young and 
old galaxy sequences as the blue and red lines, respectively.}
\end{figure}

Second, let us consider the more detailed properties of galaxies as a
function of environment. For example, Figure \ref{d4000_env} shows the
relationship between $D_n$(4000) and stellar mass as a function of
environment.  For reference, we show the rough location of each galaxy
sequence as the red and blue lines, identically in each panel of
Figure \ref{d4000_env}.  Though the position of each sequence changes
only a little with environment, how each sequence is populated changes
considerably, with old galaxies preferentially located in dense
regions, as found for these parameters by \citet{kauffmann04a}.

Other measurements of the change of galaxy properties with environment
track this same segregation of galaxy types: broad-band color and
\Sersic\ index (e.g., \citealt{hashimoto99a, blanton05b, yang07a});
H$\alpha$ emission, star-formation rate, and other spectral properties
(e.g., \citealt{lewis02a, norberg02a, gomez03a, boselli06a}); and of
course classical morphology (e.g., \citealt{dressler80a, postman84a,
guzzo97a, giuricin01a}).  It appears that among all these properties,
the controlling variables are those related to mass and star-formation
history --- not those having to do with structure
(\citealt{kauffmann04a, blanton05b, christlein05a, quintero07a}).

Figure \ref{props_env} demonstrates this conclusion. The top two rows
compare the relationship between \Sersic\ index and stellar mass for a
``young'' population and an ``old'' population, where the division is
taken to be at $D_n(4000)=1.6$. Each relationship remains remarkably
fixed as a function of density.  This means that once the stellar
population age and stellar mass are fixed, the environment of a galaxy
does not relate to its overall structure. In contrast, if we classify
galaxies according to \Sersic\ index, their stellar population ages
are a strong function of environment.

This theme persists for almost all properties of galaxies as a
function of environment. Once galaxies are appropriately classified
--- that is, by some parameter related to the star-formation history
--- then galaxy scaling laws are usually weak functions of
environment. For example, \citet{park07a} classified galaxies by color
and color gradient, and explored the scaling relationships of each
type thoroughly, finding only weak dependencies on environment.
Similarly, \citet{ball08a} found that once color was fixed, there was
no relationship between environment and their artificial neural
network determination of morphology.  Eyeball classifications of tens
of thousands of galaxies into general early and late types have been
compiled from the general public by the Galaxy Zoo project
(\citealt{bamford09a}).  They too conclude that morphology does not
depend strongly on environment once color is fixed. Furthermore, more
refined eyeball measures of spiral and ring structure and bar strength
(\citealt{vandenbergh02c, vandenbergh02b}) also appear to be weak
functions of environment.

The other striking fact in this plot is that the effect of environment
sets in at low density --- even the upper right panels of Figures
\ref{lfden} and \ref{d4000_env}, which correspond to Local Group type
environments, show more old galaxies than found in isolated regions. A
skeptic might worry that the environmental indicator used here does
not distinguish between being in a small group or at the edge of a big
cluster. However, the group-based environmental indicators of
\citet{blanton07a} show explicitly that even belonging to a small
group leads to a larger fraction of red galaxies. These results were
hinted at long ago by \citet{postman84a}, who found environmental
dependence all the way down to density enhancements similar to $N_n
\sim 1$ in our units.

A final point that the major new surveys have demonstrated about
environment is that its effects are relatively local. Much of the
environmental dependence can be understood in terms of the group
environment: what mass halo hosts the galaxy, and its position within
that halo (\citealt{blanton07a}). The larger scale density field
appears to be much less important (\citealt{kauffmann04a, blanton06a,
park07a}). The results of \citet{gomez03a}, which show a dependence of
star-formation on distance from the nearest cluster out to several
virial radii, are sometimes said to illustrate large scale effects at
work.  However, \citet{lewis02a} showed that this trend is indeed due
to group-scale effects and simply reflects the decreasing incidence of
groups as a function of distance from the cluster.

In this section, we have argued that environmental effects are
comparatively local (within the host halo), are important even at low
density, and keep galaxy scaling relationships constant with
environment.  These results suggest that the processes that transform
galaxies from one type to another are similar in all environments ---
but that they more commonly occur in dense regions.  Certainly no
effect that only acts in rich clusters can explain the observed
segregation of galaxy colors, concentrations, sizes and luminosities.

Although it is difficult to make quantitative comparisons with theory,
the observations favor an important role for so-called pre-processing
of galaxies in groups, possibly by mergers (\citealt{mihos04a}), by
gas-dynamical interactions with warm or hot gas (\citealt{fujita04a}),
or by tidal harassment (\citealt{moore98a}).  Such a picture could
explain why the segregation begins even at low densities.  In
addition, if most galaxy transformations occur on the small group
scale, then clusters may be regions that have subsumed many older
groups and thus have more than their share of ``transformed''
galaxies. This scenario would create the shift in galaxy populations
as a function of environment, without necessarily altering the scaling
relations of galaxies dramatically.

\subsection{Exceptions to environmental trends}

There are always important exceptions to the rules!  First, as we
discuss in \S\ref{sec:spirals} and \S\ref{sec:ellipticals}, there have
been recent detections of small shifts in scaling relations as a
function of environment.

Second, \citet{blanton07a} and \citet{park07a} both found a tendency
for galaxies with very close neighbors to be red or early-type at
fixed group environment.  The more detailed analysis of
\citet{park08a} revealed that galaxies within each others' virial
radii tend to have preferentially the same morphology (at fixed
environment on larger scales). \citet{weinmann06a} similarly found a
tendency towards ``conformity'' between central galaxies and their
satellites.  On small scales, galaxies appear to ``share'' information
about their type in a manner that is not explained by the larger-scale
trends.  Whether these effects are related to the star-formation
enhancements that interacting galaxies and close pairs experience
(\S\ref{mergersec}) is currently unknown.

Third, central galaxies in clusters seem to be a special case.  Though
such galaxies are early-type and similar to ellipticals, their
positions in the centers of clusters appear to affect them
substantially (\S\ref{cD}). \citet{blanton05b} found a tendency for
the most luminous galaxies (mostly central galaxies in clusters) to be
larger and more diffuse in dense regions. Such a tendency could result from
cD galaxies undergoing multiple ``dry'' mergers with other red
galaxies.

Fourth, a more subtle effect exists for central galaxies: their
properties appear to be related to the large scale density field (not
just to the host halo). For low mass host halos
($\mathcal{M}<10^{14}\mathcal{M}_\odot$), all results agree that the
central galaxy tends to be ``earlier'' type in denser regions on large
scales (e.g., \citealt{yang06a, berlind07a}). For massive host halos,
there is disagreement --- \citet{berlind07a} find that the trend
reverses, whereas \citet{yang06a} do not. It may be that the
spectroscopic classification used by \citet{yang06a}, based on a
principal component analysis (PCA) of the 2dFGRS spectra, yields
different trends than the color classification of \citet{berlind07a}.

\section{Spiral galaxies }
\label{sec:spirals}

\subsection{General description and identification}

We begin exploring the population of galaxies in more detail with the
spirals, which for our purposes we define as disk galaxies with
ongoing star-formation. In Figure \ref{spirals}, we show a sampling of
optical images of nearby spirals from the SDSS, according to the NED
classifications mentioned in \S\ref{broadband}.  We have sorted these
galaxies by absolute magnitude on the horizontal axis and $g-r$ color
on the vertical axis. An overall trend can be seen that the reddest
spirals tend preferentially to have visible dust, and are often
edge-on (\S\ref{dust}). The more luminous spirals tend (with
exceptions) to have more evident and more regular spiral structure. A
prominent feature in many spirals is a reddish, smooth, central
component, the ``bulge'' (\S\ref{bulge}).

Figure \ref{manyd-spirals} shows the distribution of spirals in the
space of absolute magnitude, color, size and \Sersic\ index.  Here we
have colored the points according to the spiral subtype: blue for Sc
and Sd galaxies, cyan for Sb galaxies, and green for Sa
galaxies. Figure \ref{manyd-spirals} (and Figure
\ref{manyd-ellipticals}, described later) represent an updated look at
the physical properties along the Hubble sequence that we first
learned from \citet{roberts94a}.  There is a strong trend for the
later type spirals to be lower in luminosity, closer to exponential,
and bluer.  The early type spirals define a (broad) red sequence, and
are concentrated and luminous.  They occupy a different place in the
size-luminosity relation, one that overlaps considerably with the
ellipticals (\S\ref{sec:ellipticals}; see also
\citealt{courteau07a}). Indeed, in these gross properties Sa galaxies
are remarkably similar to ellipticals.

The overall morphological separation according to \Sersic\ index is
driven by the relative importance of the bulge to morphological
classification.  Of course, the meaning of the \Sersic\ index for
spiral galaxies is different than for elliptical galaxies. For spiral
galaxies $n$ reflects the balance between the disk and the bulge, two
clearly distinct components. By contrast, for elliptical galaxies $n$
usually reflects the overall structure of what is apparently a single
component (\S\ref{structure}).

\subsection{Quantitative morphological measures} 
\label{morphology}

Astronomers have traditionally classified spiral galaxies from early
to late type based on three criteria: the spiral arm pitch angle, the
organization of the spiral arms, and the prominence of the bulge
(\citealt{hubble36a, sandage61b, devaucouleurs59a}).  Traditional
classifications are performed by eye from experience, but that
technique is subjective and cannot be applied to massive samples.  The
influence of the bulge criterion as reflected in the \Sersic\ index
$n$ is evident in Figure \ref{manyd-spirals}.  In contrast, the
relative importance of the spiral structure considerations vary among
classifiers. As an example, classifiers differ on at least two cases:
low bulge-to-disk ratio galaxies with little spiral structure
(\citealt{koopmann98a}), and high bulge-to-disk ratio galaxies with
late-type spiral structure (\citealt{hameed05a}). One particular
consequence of the latter ambiguity is that RC3 types (which emphasize
the bulge-to-disk ratio) tend to be earlier than those from the
Revised Shapley-Ames Catalog (\citealt{sandage81a}). These concerns
have prompted a good deal of work on the ``quantitative morphology''
of galaxies. This work has two general goals, often conflated: (1) to
reduce the complex information in galaxy images to a few simple
measurements; and (2) to classify galaxies.

Quantitative morphology often focuses on the separation of the disk
from the bulge component, often using standardized packages such as
GIM2D (\citealt{simard98a}), GALFIT (\citealt{peng02a}) or BUDDA
(\citealt{desouza04a}). We discuss the detailed nature of bulges and
bars in \S\ref{bulge}; here we discuss their demographics and
relationship with galaxy morphology. Image decomposition methods
usually treat the disk as exponential, but use more complex models for
the central component. At minimum they must include a general \Sersic\
profile bulge (they are only rarely exactly de~Vaucouleurs profiles;
\citealt{graham01a, dejong04a, laurikainen07a}).  More generally, as
\S\ref{bulge} discusses, there are at least three classes of central
galaxy components --- classical bulges (usually $n>2$), pseudobulges
(usually $n<2$; \citealt{fisher08a}), and bars (usually $n\sim
0.5$--$1$; \citealt{gadotti08c}). These components can and often do
co-exist within the same galaxy.  As a further caveat, ground-based
observations of even relatively nearby galaxies may have \Sersic\
indices artificially raised by the presence of nuclear components
(\citealt{balcells03a}).

For these reasons, three-component galaxy fits (disk, general \Sersic\
profile bulge, and bar) are now the state-of-the-art. In particular,
investigators have recognized that the bar component has a strong
influence on the fits (e.g. \citealt{dejong96a, wadadekar99a,
laurikainen05a, reese07a, gadotti08a, weinzirl08a}). Relative to
two-component analyses, the remaining bulges after bar subtraction are
smaller (often by a factor of two or more), have have lower $n$, and
are more likely to be pseudobulges.

These complications make the demographics of bulges, pseudobulges and
bars difficult to measure in large surveys. Among the larger analyses
are those of \citet{allen06a} and \citet{driver07a}, who fit
two-component models in the Millenium Galaxy Catalogue redshift
survey, and
\citet{gadotti08c}, who fit three-component models in the SDSS.
Although the stellar mass fractions of each analysis are not easily
comparable, they both conclude that classical bulges account for at
least ten times the stellar mass that pseudobulges do. As a
cautionary note, other (typically smaller but more carefully analyzed)
samples do find pseudobulges more commonly at least for low $B/T$
galaxies (\citealt{graham01a, dejong04a, laurikainen07a}).  All things
considered, probably the best current accounting of the stellar mass
fractions in various components is that of \citet{gadotti08c}, who
report that for $\mathcal{M}>10^{10}\mathcal{M}_\odot$, $32\%$ of the
stellar mass is in ellipticals, $36\%$ is in disks, $25\%$ is in
classical bulges, $3\%$ is in pseudobulges, and $4\%$ is in bars.

Beyond disk/bulge decompositions, a number of sophisticated techniques
have been developed to measure more complex ``morphological''
information.  Some methods depend on high-order decompositions of the
galaxy images, for example using Fourier modes (\citealt{odewahn02a})
or shapelets (\citealt{kelly04a}). Other approaches have used a
combination of measured parameters, for example, concentration
(\citealt{abraham94a}), asymmetry (\citealt{schade95}) and smoothness
(\citealt{conselice03c, yamauchi05a}).  More recently explored
parameters are the Gini coefficient --- a measure of the evenness of
the light distribution --- and a concentration-like parameter called
$M_{20}$ (\citealt{lotz04a, zamojski07a}).  Most investigators eschew
color information, for reasons we do not fully understand, with the
notable exception of \citet{choi07a}, who utilize galaxy color and
color gradient.

Once galaxies properties are measured, classification can be performed
using these parameters with simple cuts in parameter space (e.g.,
\citealt{choi07a}), with the aid of dimensionality reduction
techniques like principal component analysis (e.g., \citealt{ellis05a,
scarlata07b}), or using artificial neural networks (e.g.,
\citealt{storrielombardi92a, ball04a}). Doing so can effectively
classify galaxies into groups such as mergers, elliptical galaxies,
early-type spirals, late-type spirals and irregulars. In some cases
these classifications are reproducible and do a reasonable job of
reproducing classical morphological classes (e.g.,
\citealt{scarlata07b}).

None of these methods, however, distinguish well between face-on S0s,
Sa galaxies and Es --- the essential ambiguity also present in Figures
\ref{manyd-spirals} and \ref{manyd-ellipticals}. In addition,
the agreement of quantitative morphological classification with
eyeball determinations obscures the differences in approach.  The
quantitative schemes do not measure in detail the differences in
spiral structure in the later types that are essential to classical
morphology and apparent to the eyes of experts
(e.g. \citealt{elmegreen87a, vandenbergh02d}).  Generally speaking,
although quantitative measures of spiral structure exist (e.g.,
\citealt{elmegreen84a, rix95a, seigar02a, buta05a, kendall08a}), they are
rarely used for morphological classification.

Therefore, the agreement among galaxy classification systems reflects
the importance of the relatively easily measureable $B/T$ and not the
nature of the spiral structure. The latter awaits a quantitative,
objective, and large-scale analysis.

\subsection{Bulges, pseudobulges, and bars}
\label{bulge}


Spirals often have reddish, smooth central components (see Figure
\ref{spirals}). As we alluded to in \S\ref{morphology}, these
components can be divided into at least three classes: classical
bulges, pseudobulges and bars, each of which has different detailed
properties and probably different formation mechanisms
(\citealt{wyse97a, kormendy04a}). The common interpretation holds that
while classical bulges are built by mergers, pseudobulges are built
by secular processes within disks, perhaps abetted by the presence of
the bar in the bulge component.  The formation of the bars themselves
and the role of their interaction with the surrounding density field
is a matter of debate (\citealt{sellwood00a, athanassoula03a,
gadotti05a}).  We discuss all three componenets together here because
observationally they are difficult to distinguish.

The classical bulges usually but not always contain older stellar
populations than the stellar disks that surround them
(\citealt{moorthy06a}). They are often enhanced in $\alpha$-elements,
but less so than comparably luminous elliptical galaxies and with more
scatter (\citealt{ganda07a, peletier07a}).  Dynamically, many are
likely ``oblate rotators:'' consistent with having an isotropic
velocity dispersion and being flattened by rotation
(\citealt{kormendy04a}, and references therein). Finally, many but not
all have concentrated \Sersic-like profiles with high $n$. Typically,
the larger the bulge-to-disk ratio of the galaxy as a whole, the more
concentrated the bulge itself is as well (\citealt{graham01a,
gadotti08c}).

In contrast, as the review of \citet{kormendy04a} emphasizes,
pseudobulges have significant rotation support, are flatter than
classical bulges, have exponential profiles, and have circumnuclear
star-formation.  Spectroscopically, they tend to have $D_n(4000)$
similar to galaxies on the blue sequence
(\citealt{gadotti08c}). Although classical bulges lie on the
relationship between luminosity and size defined by elliptical
galaxies (the Kormendy relation), pseudobulges have lower surface
brightness and thus lower luminosity for a given size
(\citealt{gadotti08c}).  Pseudobulges are more common and more
dominant in later type (or alternatively, lower $B/T$ ratio) spiral
galaxies (\citealt{ganda06a, barazza08a}). In this manner they form a
continuous sequence of increasing bulge \Sersic\ index $n$ with
bulge-to-disk ratio (\citealt{gadotti08c}).

Although photometrically pseudobulges and classical bulges are difficult
to disentangle within the same galaxy, integral field spectroscopy
allows it more easily. For example, the recent analysis of about 40
galaxies using SAURON suggests that many bulges (of order half of
them) may actually consist of both a classical bulge and a
pseudobulge, with the pseudobulge occasionally dominating
(\citealt{ganda06a, peletier07a}).

Bars are elongated stellar structures located in the central regions
of disk galaxies \citep{kormendy04a, gadotti08b}. They are not simply
extrema in the distribution of bulge axis ratios, but rather are a
separate population (\citealt{whyte02a}). Historically, bars have been
identified by eye using a variety of (presumably subjective) criteria
\citep{devaucouleurs63a, eskridge00a}.  The most widely adopted
\emph{quantitative} technique for identifying bars is the
ellipse-fitting method, in which a bar must exhibit a characteristic
signature in both the ellipticity and position angle profiles
\citep{marinova07a, barazza08a, sheth08a}.  A simplified version of
this technique measures the difference in the axial ratio and position
angles of a best-fit ellipse to one interior and exterior isophote
\citep{whyte02a}.  In general, the visual and ellipse-fitting methods
agree about $85\%$ of the time, with egregious disagreement only
$\sim5\%$ of the time \citep{menendez07a, sheth08a}.  In edge-on
galaxies, as \citet{Combes:1981p2742} first pointed out, bars result
in ``boxy'' or ``peanut'' shaped bulges (\citealt{athanassoula05a,
bureau06a}).


About 50\%-70\% of luminous spiral galaxies have bars
(\citealt{devaucouleurs63a, eskridge00a, whyte02a, marinova07a,
menendez07a, barazza08a}).  The fractional luminosity of the bar
relative to the total light varies by more than an order-of-magnitude,
ranging from below 2\% up to 20\% \citep{elmegreen85a,
gadotti08c}. \citet{eskridge00a} found that the bar fraction in the
near-infrared was independent of morphology, while a more recent study
by \citet{barazza08a} with a considerably larger sample found that the
bar fraction is higher in blue, lower-luminosity, late-type disks
compared to more massive, red, early-type galaxies.  However, bars in
early-type galaxies tend to be stronger (i.e., more elongated) and
longer, both in an absolute sense and relative to the size of the disk
\citep{kormendy79a, elmegreen85a, erwin05a, menendez07a}.

Bars are typically dominated by evolved stellar populations
\citep{gadotti06a}, although they are also associated with enhanced
nuclear and circumnuclear star formation \citep{ho97a}, and frequently
exhibit dust lanes, which is a signature of radially inflowing gas
\citep{kormendy82a, athanassoula92a, sellwood93a, friedli93a}.  Barred
galaxies are also observed to have larger reservoirs of molecular gas
in their centers relative to unbarred galaxies
\citep{sakamoto99a, sheth05a} and flatter chemical abundance gradients
\citep{zaritsky94a, martin94a}.  

Simulations suggest that strong interactions can trigger the formation
of a bar \citep{berentzen04a}.  This hypothesis is supported by
observations showing that bars are two times more likely to be present
in ``perturbed'' galaxies (i.e., galaxies with a nearby companion),
relative to isolated galaxies \citep{varela04a}. \citet{kannappan04a}
argue similarly that blue bulges, that may be growing due to
star-formation today, are more common in close pairs.  However, we
found no publications describing a large-scale exploration of whether
the detailed nature of the bulges or the presence of a bar is
otherwise strongly related to environment.

\subsection{Atomic and molecular gas content}
\label{gas}

Most of the baryons in the Universe appear to be in warm or hot gas in
the space between galaxies (in groups and clusters, this gas is
directly detectable in the X-rays; \citealt{mulchaey00a,
  rosati02a}). According to
the census of \citet{read05a}, about $\sim 80\%$ of the remaining
baryons are in stars, and the rest are in atomic gas ($\sim 10\%$) and
molecular clouds ($\sim 10\%$ ). The cold gas is overwhelmingly
located in galaxies with gas disks (though not completely absent in
ellipticals; \S\ref{coldgas}).  In particular, luminous spirals tend
to have about 10\%--20\% 
of their baryonic content in the form of neutral
hydrogen and molecular clouds.

The census of atomic hydrogen in galaxies relies on the 21-cm
hyperfine transition, either in blind surveys such as HIPASS and
ALFALFA, or in compilations such as those of
\citet{springob05a}. THINGS has recently released a survey of 34
nearby galaxies with extremely detailed observations
(\citealt{Walter:2008p2836}).  As an illustration, the top panel of
Figure \ref{gas_props} shows the ratio of atomic to stellar gas mass,
as a function of stellar mass.  The greyscale shows the results from
\citet{springob05a} (using galaxies for which we have the stellar mass
from SDSS). The overlaid points are from \HI\ observations of SINGS
galaxies \citep{kennicutt03a, draine07a, walter07a, leroy08a}.
Clearly low mass galaxies tend to have much higher atomic gas content
than massive galaxies, as has been known for a long time
(\citealt{young91a}).

Determinations of molecular hydrogen mass in galaxies typically rely
on carbon monoxide as a tracer, using the CO ($1\rightarrow0$)
transition at 2.6 mm (\citealt{young95a}).
In the middle panel of Figure \ref{gas_props}, we show the ratio of
H$_2$ to atomic hydrogen content for the SINGS galaxies. Although for
luminous galaxies a substantial fraction of the gas, indeed often the
majority, is in molecular form, the fraction appears to decline at
lower luminosities.  The bottom panel of Figure \ref{gas_props} shows
the total gas fraction for SINGS galaxies as a function of stellar
mass. Owing mainly to a fractional increase in atomic hydrogen, low
mass galaxies are much more gas rich --- they have been less efficient
at turning gas into stars than their massive counterparts.



A class of spiral galaxies exists with very little atomic gas; they
are known as ``anemic'' or ``passive'' spirals
(\citealt{vandenbergh76a, elmegreen02a}).  For spiral galaxies of a
given optical diameter and morphological type, those in clusters to be
``\HI-deficient'' relative to those in the field, and to have a
correspondingly smaller \HI\ radius as well, and are thus more likely
to be anemic (e.g., \citealt{giovanelli83a, haynes84a, warmels88a,
cayatte94a, vogt04a, boselli06a, levy07a, chung08a}). Clearly the
atomic hydrogen in galaxies is being affected by the cluster
environment, likely by ram pressure stripping or related processes
(\citealt{gunn72a}) --- often seen in the process of happening,
e.g. by \citet{Chung:2007p2771} --- and at least in some cases is
resulting in the stripping of atomic gas from the outside in.
Meanwhile, the molecular gas content is not deficient, at least when
based on CO tracers (\citealt{Kenney:1988p3425, vandenbergh91a,
boselli94a, casoli98a, boselli06a}).  The high molecular clouds
densities may prevent their stripping.

Despite its importance, ram pressure stripping in rich clusters cannot
explain the broader segregation of galaxy types, which extends even to
small groups (\S\ref{environs}).  In those regions, which do exhibit
environmental effects according to galaxy color, most analyses
nevertheless have found few H~{\sc i} deficient galaxies
(\citealt{stevens04a, kilborn05a}).  In small groups, mergers or other
processes might be lowering the \HI\ content while simultaneously
altering the galaxy morphology, which would not necessarily result in
deficiencies as they are currently defined.


\subsection{Star formation}
\label{sfr}

A variety of both direct and indirect techniques have been developed
to measure the globally averaged star-formation rates (SFRs) of
galaxies, as reviewed in detail by \citet{kennicutt98a} (see also
\citealt{hopkins03a, moustakas06b, calzetti08a}).  More recently,
observations with \emph{Spitzer} and \emph{GALEX} have led to a new
class of hybrid SFR indicators that use a weighted combination of
UV/optical and infrared luminosity to minimize the systematic effects
of dust obscuration (\citealt{gordon00a, kennicutt07a, calzetti07a,
  rieke08a}.

Surveys like the SDSS and \emph{GALEX} have also quantified the
star-formation histories of galaxies in unprecedented detail (e.g,
\citealt{gallazzi06a, panter07a, cidfernandes07a}).  One method of
parameterizing the star-formation history is to measure the birthrate
parameter $b=\mathrm{SFR}/\avg{\mathrm{SFR}}$, or the ratio of the
current rate to the past-averaged rate of star formation
\citep{kennicutt83b, kennicutt94a, boselli01a, brinchmann04a}.  The
birthrate parameter is also proportional to the \emph{specific} SFR,
or the SFR relative to the present-day stellar mass.  Analyses using
H$\alpha$ (\citealt{brinchmann04a}) and UV measurements
(\citealt{salim07a, schiminovich07a}) have shown that $b$ is a slowly
declining function of stellar mass ($b\propto M^{-1/3}$).  Most
galaxies are in a relatively ``quiescent'' state, with only about 20\%
in star-bursting systems ($b>2$; \citealt{kennicutt05a}).  On this
``star-forming'' sequence, the surface density of star-formation in
galaxies is nearly constant.  This relationship is truncated where
massive ellipticals begin to dominate the mass function, as their SFRs
are often unmeasurably small using these techniques.

Comparison of the local gas surface density with the SFR surface
density reveals the relationship first discussed by
\citet{schmidt59a}: $\Sigma_{\mathrm{SFR}} \propto
\Sigma_{\mathrm{gas}}^n$, with $n\sim 1$--$2$ (\citealt{kennicutt89a,
  kennicutt98c, wong02a, boissier07a}). Recent results suggest a much
more direct relationship between SFR and molecular gas content
(\citealt{blitz06a, kennicutt07a, bigiel08a, leroy08a}). The anemic
spiral galaxies with low atomic and molecular gas content
(\S\ref{gas}) show correspondingly low SFRs, probably owing to their
low gas surface density (\citealt{elmegreen02a}).  Interestingly, the
efficiency of the conversion of molecular gas into stars is nearly
independent of the galaxy type, its larger-scale environment, or the
particular local conditions within the galaxy (\citealt{rownd99a,
leroy08a}).

Resolved H$\alpha$ measurements imply a threshold gas density defined
by local conditions, below which the SFR is much lower than predicted
by the Schmidt law (\citealt{kennicutt89a, martin01a}). Typically, the
gas density falls below the threshold in the outer
disk. Theoretically, such a threshold could be imposed by the
\citet{toomre64a} stability criterion, which predicts the conditions
under which disks are stable to collapse and star formation
(\citealt{schaye04a}).  In contrast to the H$\alpha$ results, UV
observations from GALEX reveal a much smoother transition at large
disk radii and few direct signs of a stability threshold
(\citealt{boissier07a, leroy08a}).

As discussed in \S\ref{environs}, the fraction of galaxies with young
stellar populations is a strong function of environment.  However, it
appears that the details of the star-formation history are not closely
related to environment: in fact, long time-scale stellar population
indicators such as $D_n$(4000) and $g-r$ are sufficient to describe
the dependence.  Shorter time-scale indicators like H$\delta$
absorption (\citealt{kauffmann04a}) and H$\alpha$ emission
(\citealt{cooper08a}) are related to $D_n$(4000) and $g-r$ in a manner
that is independent of environment.  This result may imply that the
time scale for any shut-off of star-formation in dense regions is
usually rather long for most galaxies.  

Spatially resolved studies of spiral galaxies in clusters suggest that
their star formation is most significantly reduced in their outer
disks, similar to the results on H{\sc i} deficiency
(\citealt{koopmann98a, vogt04a, koopmann06a}). This effect is likely a
further sign of the impact of ram pressure stripping in those regions.

\subsection{Dust content}
\label{dust}

A small fraction, about 0.1\%, of the baryonic mass in spiral galaxies
is in the form of dust, but its presence is disproportionately
important to the evolution of galaxies and our observations of
them. \citet{draine03a} reviews the properties of dust in galaxies,
which consists primarily of grains of graphite or silicon a micron or
less in size, with between 1\%--5\% of its mass in polycyclic aromatic
hydrocarbons (PAHs).  Most of the interstellar Mg, Si, and Fe, and
much of the carbon, is in the form of dust grains. \citet{draine07a}
has performed a recent census of the dust content of galaxies using
the SINGS sample (see also \citealt{dale07a}).

Dust tends to efficiently absorb the UV/optical light emitted by the
massive stars produced by recent star-formation.  Since the dust
temperatures are of order 10--100 K, this light is then reradiated in
the infrared (\citealt{obric06a}).  Especially in the galaxies with
the greatest star-formation rates, the UV/optical light therefore
traces only the unobscured star-formation, and IR observations are
necessary to determine the rest (\S\ref{sfr}).


The optical dust extinction also of course affects the optical
emission and the broad-band colors in the near-infrared and bluer. For
galaxy observations, this extinction can be a confusing factor,
especially since its impact is highly dependent on the inclination
angle of the galaxy disk relative to the line-of-sight (as well as
other geometrical effects; \citealt{witt00a}). Thus, the colors of
spirals galaxies are a strong function of their axis ratios, an effect
that sometimes needs to be accounted for (e.g., \citealt{maller08a}
and references therein).

Like the molecular gas content, the dust-to-gas ratio of galaxies
appears to be a strong function of galaxy mass. For example,
inclination-dependent reddening is substantially weaker for low mass
galaxies (e.g., \citealt{tully98a}, \citealt{maller08a}).  Using a
sample of edge-on galaxies, \citet{dalcanton04a} claim that galaxies
with $V_c<100$ km s$^{-1}$ have no dust lanes. In addition, the low
mass galaxies tend to show less dust emission relative to stellar
emission, and fewer PAHs relative to dust as a whole
(\citealt{draine07a}). That low-mass galaxies tend to be less dusty is
likely related to their relatively low gas-phase metallicities
(\S\ref{yeff}).

\subsection{Chemical history}
\label{yeff}

The gas phase chemistry of spiral galaxies also shows some revealing
trends, as Figure \ref{manyd-spectro} demonstrates.  As shown by
\citet{tremonti04a}, mass and metallicity are strongly correlated at
low masses, with the metallicities approaching an approximately
constant value at masses $\mathcal{M}\simgreat
10^{10}\mathcal{M}_\odot$.

A more revealing measurement than gas phase metallicity is often the
{\it effective yield}: the metallicity relative to the gas
fraction. In closed boxed chemical evolution models, the metallicity
tracks the gas fraction as $Z= y_t \ln f_{\mathrm{gas}}^{-1}$, where
$y_t$ is the nucleosynthetic yield, the fraction of metals that are
returned to the interstellar medium by stars. The effective yield is
defined as:
\begin{equation}
y_{\mathrm{eff}} = \frac{Z}{\ln f_{\mathrm{gas}}^{-1}}, 
\end{equation}
The ratio $y_{\mathrm{eff}}/y_t$ then is simply the metallicity
relative to a closed box system with the same gas mass fraction.  That
is, although the metallicity can be low simply because there has been not
much star-formation, if the effective yield is low there must have
been some violation of the closed box model (\citealt{tinsley80a,
pagel97a}).  

Estimates of the effective yield by \citet{tremonti04a} and more
direct measurements by \citet{garnett02a} and \citet{pilyugin04a}
indicate that $y_{\mathrm{eff}}$ increases with circular velocity and
stellar mass.  Assuming that $y_t$ is not a strong function of galaxy
mass (as could be the case if the IMF is variable;
\citealt{koppen07a}), this trend implies for low mass galaxies either
an inflow of pristine gas or a metal-enriched outflow
\citep{martin02a, veilleux05a}.  However, a detailed analysis by
\citet{dalcanton07a} demonstrates that inflow of pristine gas can only
reduce $y_{\mathrm{eff}}$ by a limited amount (30\%--50\% for
reasonably gas-rich galaxies) and cannot explain the observed trends.
Metal-enriched outflows, on the other hand, are both physically
motivated (\citealt{maclow99a}) and can significantly reduce
$y_{\mathrm{eff}}$. Although a small but metal-rich outflow may
explain the trends, there is no necessity for the ejected fraction of
gas to exceed that in high-mass galaxies.  In fact, most evidence
suggests that low mass systems have the same baryonic fraction as high
mass systems and thus have experienced the same fractional outflow
(\citealt{blanton07b}).

Interestingly, the mass-metallicity relation seems to be a strong
function of $r_{50}$, with the smallest galaxies at a given stellar
mass having the highest metallicities (\citealt{ellison08a}). These
researchers conclude that the most likely explanation is that galaxies
with high stellar mass surface densities are those that have most
efficiently converted gas to stars. If pristine gas is not remixing
from the outer radii of the galaxy, then such galaxies will have lower
gas fractions locally and thus higher metallicities.  In this context,
it is tempting to note that the smallest spirals at a given stellar
mass are the early type spirals, a possibly related fact.

This mass-metallicity relationship for spiral galaxies appears to be
only a weak function of galaxy environment (\citealt{mouhcine07a,
  cooper08a}).  There are 0.02 dex trends with respect to environment,
though these are small relative to the overall distribution of
metallicities. According to \citet{cooper08a}, the noise in their
environmental indicators may be washing out a stronger existing
relationship, and these trends might explain about 15\% of the scatter
in the mass-metallicity relationship.

\subsection{Disk edges and extended galactic disks}
\label{xuv}

The optical surface brightness profiles of galaxy disks often exhibit
a sharp edge, around 3--5 times the exponential scale length
(\citealt{vanderkruit01a}).  Typically, at the edge the surface
brightness dips below 25--26 mag arcsec$^{-2}$ in $B$, motivating
classical parameters such as $R_{25}$ (the radius at which $\mu_B =
25$ mag arcsec$^{-2}$; \citealt{devaucouleurs91a}) and the
\citet{holmberg58a} radius (where $\mu_B = 26.5$ mag arcsec$^{-2}$).

However, it has recently been recognized that for many disk galaxies
this truncation is not complete (\citealt{pohlen04a}). The census of
\citet{pohlen06a} finds that while about 56\% exhibit a downward break
(either a classical sharp break or one described by a steeper
exponential profile), about 24\% exhibit a shallower exponential
profile on the outside (e.g. \citealt{erwin05a}), and 10\% exhibit no
measurable break. Apparently the latter category includes some
galaxies that extend up to 10 scale radii (e.g., NGC 300;
\citealt{blandhawthorn05a}). \citet{pohlen06a} found that the
remaining 10\% of their sample were more complicated and consisted of
a mix of breaks.

Meanwhile, the neutral gas in galaxies as seen at 21-cm tends to
extend considerably further than the disk break, typically by a factor
of two (see the recent compilation by the Westerbork \HI\ survey;
\citealt{swaters02a}). Therefore, the break in
optical light could relate to the similar breaks implied by H$\alpha$
measurements of spatially resolved star-formation (\S\ref{sfr}).

It has recently become clear that even beyond the edges of stellar
disks star-formation can still occur, which may explain why some
galaxies have no downward break or edge. \citet{thilker07a} has
recently reported from \emph{GALEX} imaging that over 20\% of spiral
galaxies have significant UV emission in the outer disk. As one might
expect, the H{\sc ii} regions associated with the UV emission are low
metallicity, typically 10\%--20\% solar
(\citealt{gildepaz07a}). Galaxies with such disks tend to be gas-rich
relative to other spirals at the same luminosity, and overdensities in
the gas distribution tend to correlate with the UV light. Similarly,
\citet{christlein08a} detect H$\alpha$ up to $1.5\times R_{25}$ on
average and up to $2\times R_{25}$ in some galaxies. Such galaxies
consist of rotating disks with flat rotation curves. Both of these
sets of results indicate that some star-formation can occur at large
radii where most of the gas is stable against collapse, probably due
to local disturbances and overdensities.

\subsection{Tully-Fisher relation}
\label{tully-fisher}

The gas and (thin-disk) stars in spiral disks orbit coherently in
nearly circular orbits around the galactic center: they are
``rotation-supported'' objects.  At high luminosity at least, spiral
galaxies often have flat or peaked rotation curves, allowing one to
define a maximum circular velocity $V_c$ or an asymptotic circular
velocity $V_a$, which are often but not always similar.  The most
commonly used tracers of the outer disk dynamics are H$\alpha$ and
H~{\sc i} emission, with the latter generally detectable at larger
galactic radii than the former. The full details of measuring the
internal rotation curves were recently reviewed by \citet{sofue01a},
and the mass-modeling of these data are the subject of considerable
debate (e.g., most recently
\citealt{barnes04b, dutton05a, kassin06a, spano08a}). We 
concentrate here on the global dynamics.

According to the well-established \citet{tully77a} relation (TF),
galaxy luminosity $L$ is related to $V_c$ as a power-law $V_c \propto
L^{\alpha}$ at high luminosities, with some scatter.  Figure \ref{tf}
illustrates the $I$-band TF relation for luminous galaxies from
\citet{courteau07a}. In virtually all analyses, an attempt is made to
correct for internal extinction within each galaxy, which in general
is a complex function of type, luminosity, and inclination
(\citealt{tully98a, maller08a}).  The slope of the TF relationship is
usually determined using a linear regression, either of $\log_{10}
V_c$ on $\log_{10} L$ (the ``inverse'' TF relation) or vice-versa (the
``forward'' TF relation).

Because of the scatter, the inverse and forward TF relations do not
have slopes that are simply inverses of each other --- $\alpha$ will
be larger if the forward fit is performed, by an amount that depends
on the selection effects and error distribution of the
sample. \citet{courteau07a} instead use a bisector method that
accounts for uncertainties in both variables. We do not advocate any
one method; in practice, the correct choice depends on the error
distributions, the sample selection effects, and the desired
goals. The most complete discussion of these issues remains
\citet{strauss95a}, but see \citet{masters06a} for a more modern
view. We simply point out that the TF relations derived from different
groups differ due to such choices.

Recent compilations (\citealt{verheijen01b, kannappan02a, masters06a,
pizagno07a, courteau07a}) find $\alpha \sim 0.27$--$0.35$, with a
scatter equivalent to about $0.15$--$0.4$ mag in the luminosity
direction. The low scatter holds for \citet{courteau07a} and other
samples that were selected for simplicity and for best
distance-measure performance.  The higher scatter value from
\citet{pizagno07a} occurs because they use a wider range of
morphological types, in particular including ``peculiar'' systems and
barred galaxies. The TF slope is a function of bandpass, with $\alpha$
decreasing towards longer wavelengths, as one might expect as spirals
become redder with increasing mass (cf. Figure \ref{manyd-spirals}).

The residuals from the TF relation in red bands (such as SDSS $i$ or
Bessell $I$) usually appear to be only weakly related to
any other properties, suggesting they are dominated by the dark matter
to stellar mass ratio variation. For bluer bands like $g$ there are
residuals associated with color, expected since in those bands the
stellar mass to light ratio varies with color (\citealt{bell01b,
  pizagno07a}). In red bands, the strongest residuals appear to be the
tendency for earlier-type (\citealt{masters06a, courteau07a}) or more
concentrated or redder spirals (\citealt{kannappan02a, pizagno07a}) to
have higher $\alpha$.  In particular, at $V_c\sim 250$ km s$^{-1}$, Sa
galaxies are less luminous than later types by roughly 0.5 mag.

Although the standard TF relation is appropriate for massive galaxies,
for less massive galaxies, for example where $V_c<100$~km~s$^{-1}$, it
has considerably more scatter and deviates from a power law ---
galaxies at a given circular velocity are less luminous than the high
luminosity TF would predict (e.g., \citealt{mcgaugh05a}).  As we saw
above, at these scales the disk gas mass starts to become dominated by
the gas contribution. By including the neutral and molecular gas mass
to define a total baryonic mass, \citet{mcgaugh05a} and
\citet{begum08a} claim that the TF power law continues even to these
scales. This change probably reflects a decrease in the time-averaged
efficiency of turning baryonic matter into stars in lower mass
galaxies.

These dynamical relationships appear to be at best a weak function of
environment for luminous galaxies. For example, in a sample of 165
galaxies, \citet{pizagno07a} found no evidence of environmental
dependence.  
A hint of dependence is seen in the lower luminosity sample of
\citet{blanton07b} at $V_c < 70$ km s$^{-1}$.
 
\section{Lenticulars}
\label{sec:lenticulars}

The lenticular galaxies, or S0s, are classified in the Hubble sequence
in between the spiral population and the ellipticals.  They are disk
galaxies, but like ellipticals are smooth, concentrated, and have low
specific star-formation rates (\citealt{caldwell93a}).  They are
distinct from the anemic spirals (\S\ref{sfr}), in that they have very
little molecular gas or spiral structure.  However, some S0 galaxies
might form from anemic spirals whose spiral wave pattern has
disappeared due to its short-lived nature (\citealt{elmegreen02a}),
possibly aided by tidal ``harassment'' in dense regions
(\citealt{moore98a}).

Figure \ref{lenticulars} shows some images of typical lenticular
galaxies as classified by NED. For this figure, we rejected through
visual inspection about one-quarter of the NED classifications as
being clearly incorrect (ambiguous cases were kept). The broad-band
properties of S0s are shown as the orange points in Figure
\ref{manyd-ellipticals}. Evidently, in these properties they are
practically inseparable from ellipticals.

Consequently, though there is a clear physical distinction between
Es and S0s, few researchers have implemented objective measurements on
large data sets that neatly separate the two populations. Visual
classification schemes usually rely on the axis ratio, which for
inclined S0s makes their disk-like nature clear, or the strong break in
surface brightness at their disk edge (similar to that of spirals;
\S\ref{xuv}). However, S0s with a ring or
multiple rings are also obvious visually, and barred S0s can exhibit
the upturn in radial profile discussed in \S\ref{xuv}.  S0s are of
particular interest because they appear to be structurally similar to
spirals, but to have ended their star-formation.

Given their observed properties, the most compelling question about S0s is
whether they are spirals that ran out of gas and faded onto the red
sequence. One simple way of testing this hypothesis is to ask whether
they scatter to low luminosities in the Tully-Fisher relation, as we
would expect for a faded population.  Our understanding of the
Tully-Fisher relation is poorer for S0s than spirals because S0s lack
significant H$\alpha$ or 21-cm emission, making dynamical mass
estimates challenging.  Nevertheless, beginning with
\citet{dressler83b}, a number of investigators have tried to measure
S0 dynamics using stellar absorption lines, which generally yield lower
signal-to-noise velocities. \citet{bedregal06a} compile a set of
dynamically analyzed S0 measurements and find that the S0 Tully-Fisher
relation is offset in the $K$-band from that of spirals by about 1 mag
(see also \citealt{hinz03a}). In Figure \ref{tf}, we show these S0
galaxies in the $I$ band, where the offset is closer to about 1.5 mag
relative to the spirals in \citet{courteau07a}; naturally, the actual
offset depends on the Tully-Fisher zeropoint and is somewhat
uncertain.

This offset is about what one would expect if S0s were simply
``faded'' versions of spiral galaxies, whose star-formation had
shut-off several billion years ago, but were otherwise well
represented by the early-type spiral population today. In this manner,
the S0s may form a continuum with the Sa galaxies or other red
spirals, which as noted in \S\ref{tully-fisher} are also offset from
Tully-Fisher by about 0.5 mag (at high $V_c$ at least). Similar
conclusions result from the study of globular cluster populations,
which are substantially more frequent per unit luminosity in S0s than
in spirals of similar mass; this trend would result if the stellar
population faded while the globular cluster population remained
unchanged (\citealt{barr07a}).

However, one might na\"ively expect that if S0s are merely dead
spirals then the luminosity and surface brightness distributions of
S0s would be systematically fainter than that of spirals, at least
within a common environment.  A recent analysis of \citet{burstein05a}
instead showed that S0s tend to be brighter than any other spiral type
even at fixed environment. Similarly,
\citet{Sandage:2005p1910} shows that the typical surface brightness of
S0s is larger than that of spirals. Similar trends can be
seen in Figures \ref{manyd-spirals} and \ref{manyd-ellipticals}.  
In more detail, S0s have larger bulge-to-disk ratios than could result
from fading disks in early-type spirals (\citealt{dressler80a,
christlein04a}). As a cautionary note, analyses that
account for the bar and a general \Sersic\ profile for the bulge
indicate much smaller bulges for S0s and might change this conclusion
(\citealt{laurikainen05a}).

These latter facts support a hypothesis that mergers (which would
increase the overall stellar mass) are responsible for the
transformation of S0s. If mergers sparked globular cluster formation
they might also account for the increased globular cluster frequency.
However, merger scenarios have a hard time explaining the S0
Tully-Fisher relation unless the progenitor population was quite
different from any that exists in abundance today.

Interestingly, the dependence of S0 fraction and S0 properties on
environment has not been studied with the new, large samples
available, in large part due to the difficulty in identifying them
automatically or unambiguously. Thus, while \citet{dressler97a} show
that S0s become relatively more frequent as one approaches the centers
of clusters, no significant improvement or refinement of that
measurement has been undertaken for the nearby Universe.

\section{Ellipticals}
\label{sec:ellipticals}

\subsection{General description and identification}

Elliptical galaxies are recognizable by their smooth, symmetric, and
deceptively simple-looking appearance. Figure \ref{ellips} shows some
typical elliptical galaxies drawn from the SDSS, using the
classifications reported by NED.  In this case, we had to reject about
one-third of the classifications as obviously incorrect.  Figure
\ref{manyd-ellipticals} shows the distribution of their broad-band
properties along with the S0s. They are concentrated, uniformly red,
and follow a reasonably tight size-luminosity relation.

The past two decades of research, however, have revealed the
complexity of their dynamical structure, their star-formation history,
and their assembly history. In general terms they are dynamically
supported by velocity dispersion, but often with significant
rotational support as well. Their stellar populations are old, ongoing
star-formation is rare, and their cold gas content is low (around or
less than 1\%). However, there are numerous indications that at least
some ellipticals were assembled late, after the hey-day of their
star-formation. In addition, the small amount of star-formation that
does occur in ellipticals seems to be similar in nature to that in
spirals: molecular clouds in gas disks forming stars, at the same
(low) efficiency found in spirals.

Among luminous ellipticals there are two discernible classes, those
with and without cores (\S\ref{core}; \citealt{Kormendy:2008p1880} and
references therein). The ellipticals with nuclear cores, relative to
those without, tend to be more luminous, have $b/a$ closer to unity,
have boxier isophotes, less rotational support, and more signs of
triaxiality.

Among lower luminosity ellipticals there are at least three classes:
low surface brightness, exponential galaxies typed as dE or sometimes
Sph (such as NGC 205); high surface brightness, concentrated galaxies
usually typed as cE (such as M32); and ultra-compact dwarfs with
$r_{50}$ as small as 10 pc
(\citealt{Drinkwater:2004p2799}). \citet{Kormendy:2008p1880} reviews
the distinction between the first two classes, arguing that the cE
population is most similar to the giant elliptical
population. \citet{ferrarese06a} presents the alternative argument for
dE galaxies.

Relative to S0s, the surface brightness profiles of ellipticals appear
to have no ``edge,'' as S0s do. However, in their other properties ---
sizes, profile shapes, colors, symmetry, and smoothness --- Es and S0s
are very similar.  Thus, the recent literature on very large samples
is plagued with ``elliptical'' or ``early-type'' samples that actually
include a fair number of S0s as well as early-type or edge-on
spirals. Interestingly, these interlopers do not appear to affect many
of the scaling relations we discuss here; nevertheless, they can be
important in some circumstances and we emphasize again the importance
of separating these two populations.

\subsection{Structural trends}
\label{structure}

\begin{figure}
\plotone{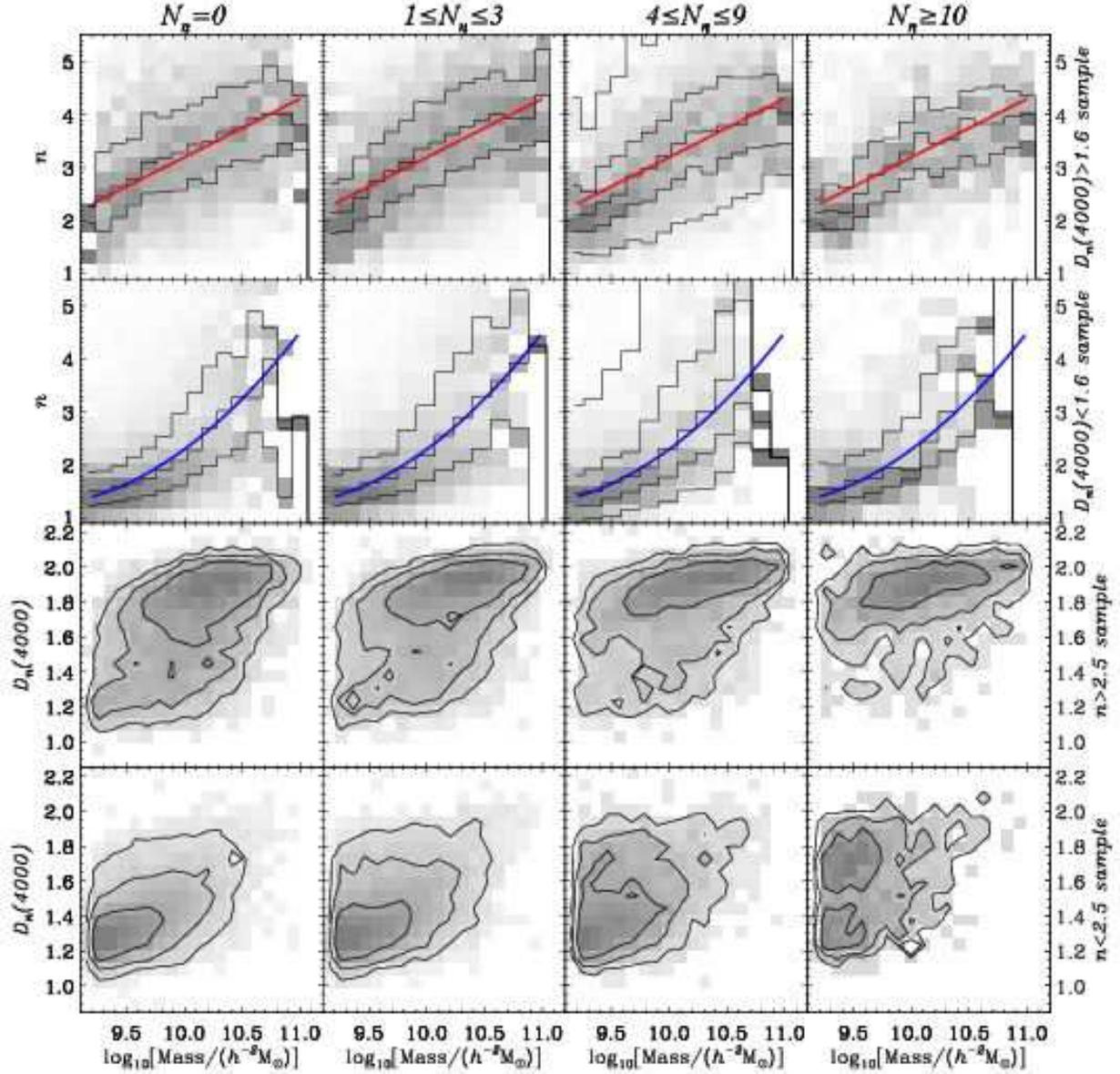}
\caption{\label{props_env} Demonstration of the asymmetric relations
between structure and star-formation history. Each column isolates a
range of densities using the number of neighbors $N_n$, as defined
in \S\ref{environs}. The top two rows show the relationship
between \Sersic\ index and stellar mass in an ``old'' sample and a
``young'' sample. For the old sample plots, the red line remains the
same in all plots.  For the young sample plots, the blue line remains
the same. The bottom two rows show the relationship between
$D_n(4000)$ and stellar mass in a ``concentrated'' sample and a
``diffuse'' sample. At fixed \Sersic\ index the stellar population
ages clearly depend strongly on density (even at fixed mass).
However, at fixed stellar population age, the \Sersic\ index does not
vary significantly with density.}
\end{figure}

\begin{figure}
\plotone{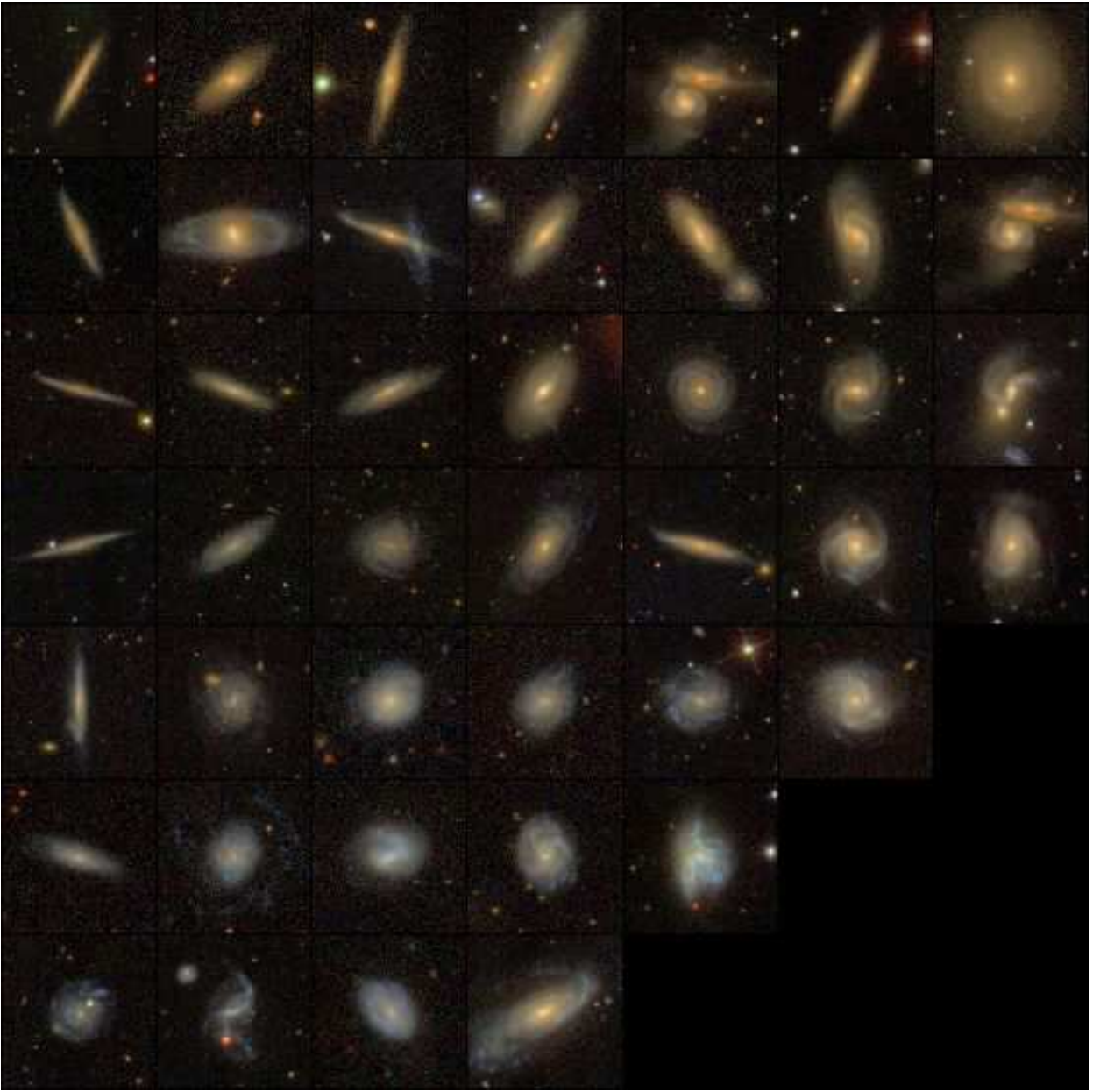}
\caption{\label{spirals} SDSS images of spiral galaxies, selected 
according to classifications in NED to be Sa--Sd (including barred
types). The images are sorted by absolute magnitude in the horizontal
direction, ranging between $M_r -5\log_{10} h\sim -18.5$ and $-22$
from left to right, and $g-r$ color in the vertical direction, ranging
between $0.2$ and $0.9$~mag from the bottom to the top.  Thus, the
brightest, reddest spirals are in the upper-right.  The galaxies shown
were selected randomly, except we excluded two cases from the original
set due to image defects.}
\end{figure}

\begin{figure}
\plotone{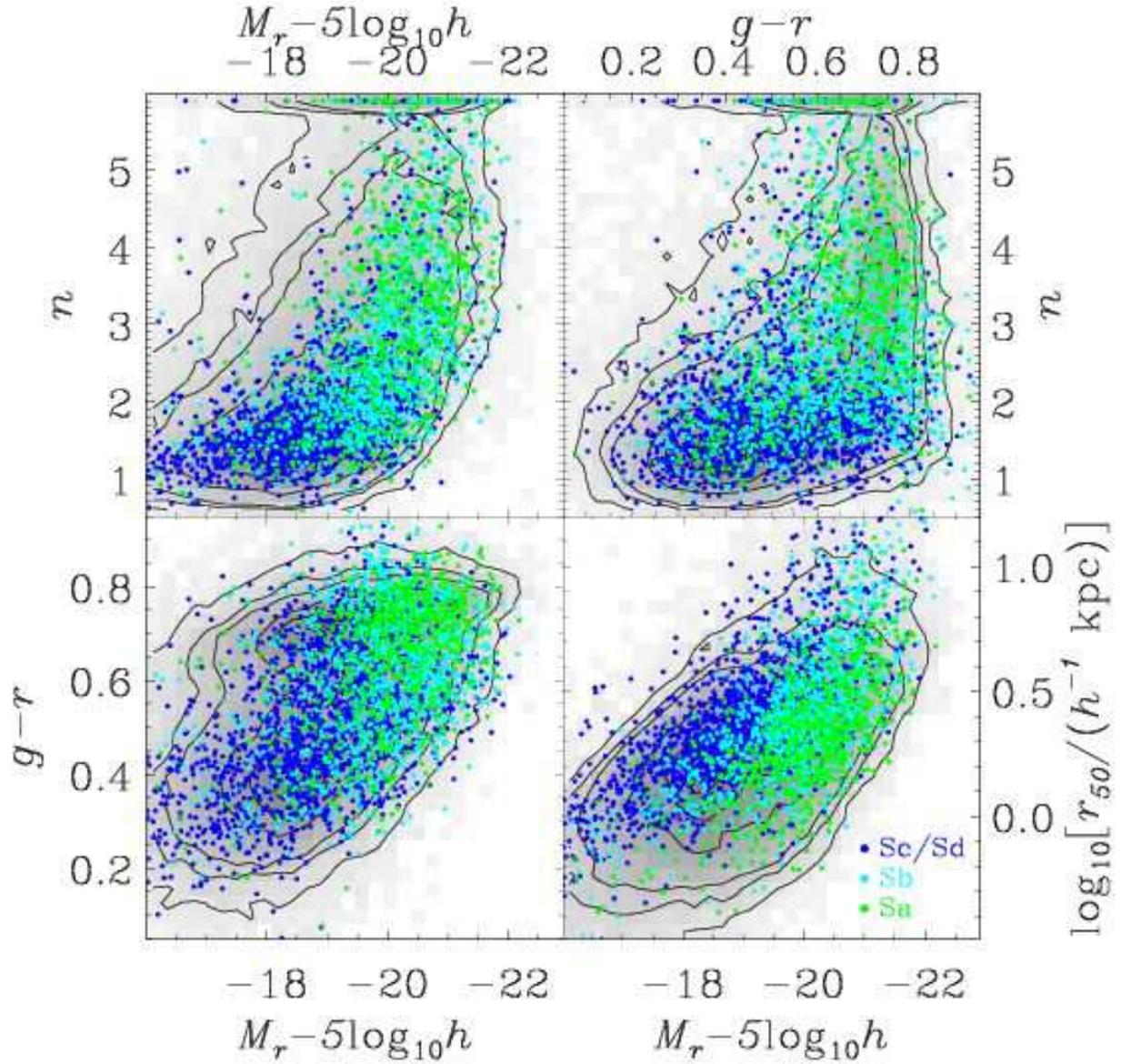}
\caption{\label{manyd-spirals} Distribution of various spiral galaxy
types in optical broadband properties. The underlying greyscale and
contours are from the full data of Figure~\ref{manyd}. Using
classifications stored in NED, we overplot the positions of Sa
(green), Sb (cyan), and Sc or Sd (blue) galaxies. We include the
barred varieties. }
\end{figure}

\begin{figure}
\plotone{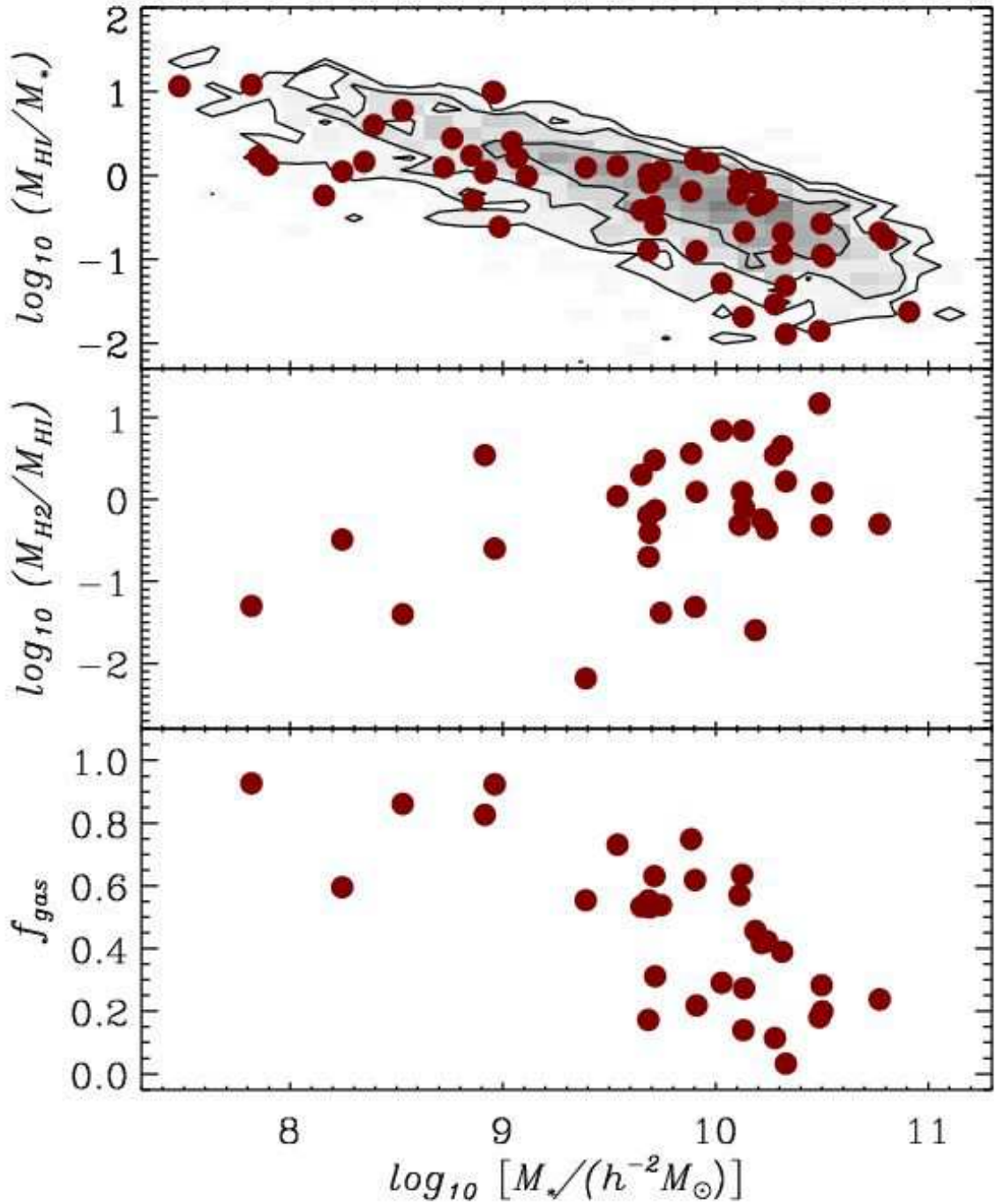}
\caption{\label{gas_props}  Neutral and molecular gas content 
of spiral galaxies. The top panel shows the fraction of neutral gas
relative to the stellar mass, as a function of stellar mass. The
greyscale is from a combination of data in the
\citet{springob05a} compilation of HI data, matched to the SDSS sample
(using the formulae of \citealt{bell03a} to determine stellar
mass). The points are from SINGS.  The middle panel shows the ratio of
molecular hydrogen gas mass (inferred from the CO (1$\rightarrow$0)
transition) to neutral gas mass for the SINGS sample.  The bottom
panel shows the total neutral plus molecular gas fraction as a
function of stellar mass.}
\end{figure}

\begin{figure}
\plotone{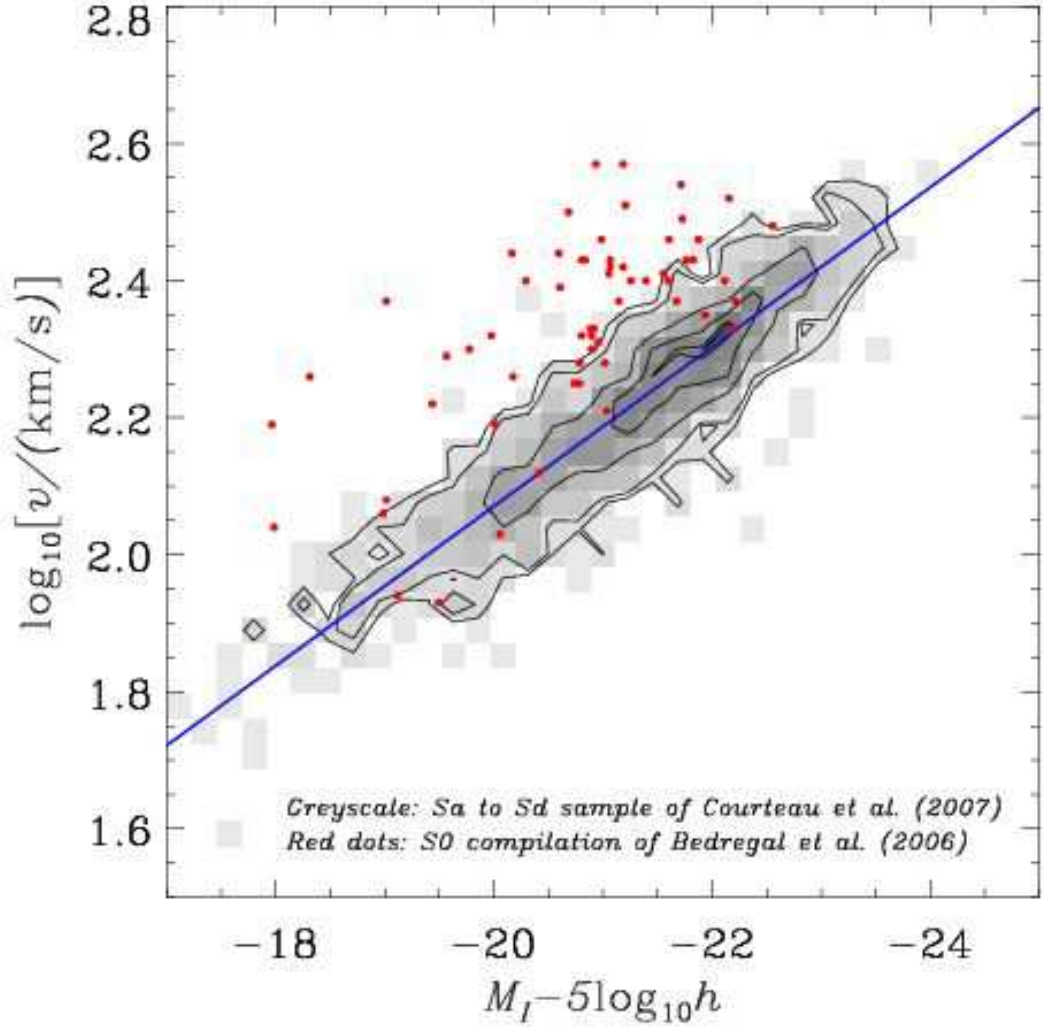}
\caption{\label{tf} Tully-Fisher relation in the $I$-band
(Vega-relative). The greyscale and contours show the data from
\citet{courteau07a}, along with the fit to all morphological types
given in their Table~2. The red points are the S0 data from the
compilation of \citet{bedregal06a}, converted to $I$ band using the
approximate relation $I=K+1.99$.}
\end{figure}

\begin{figure}
\plotone{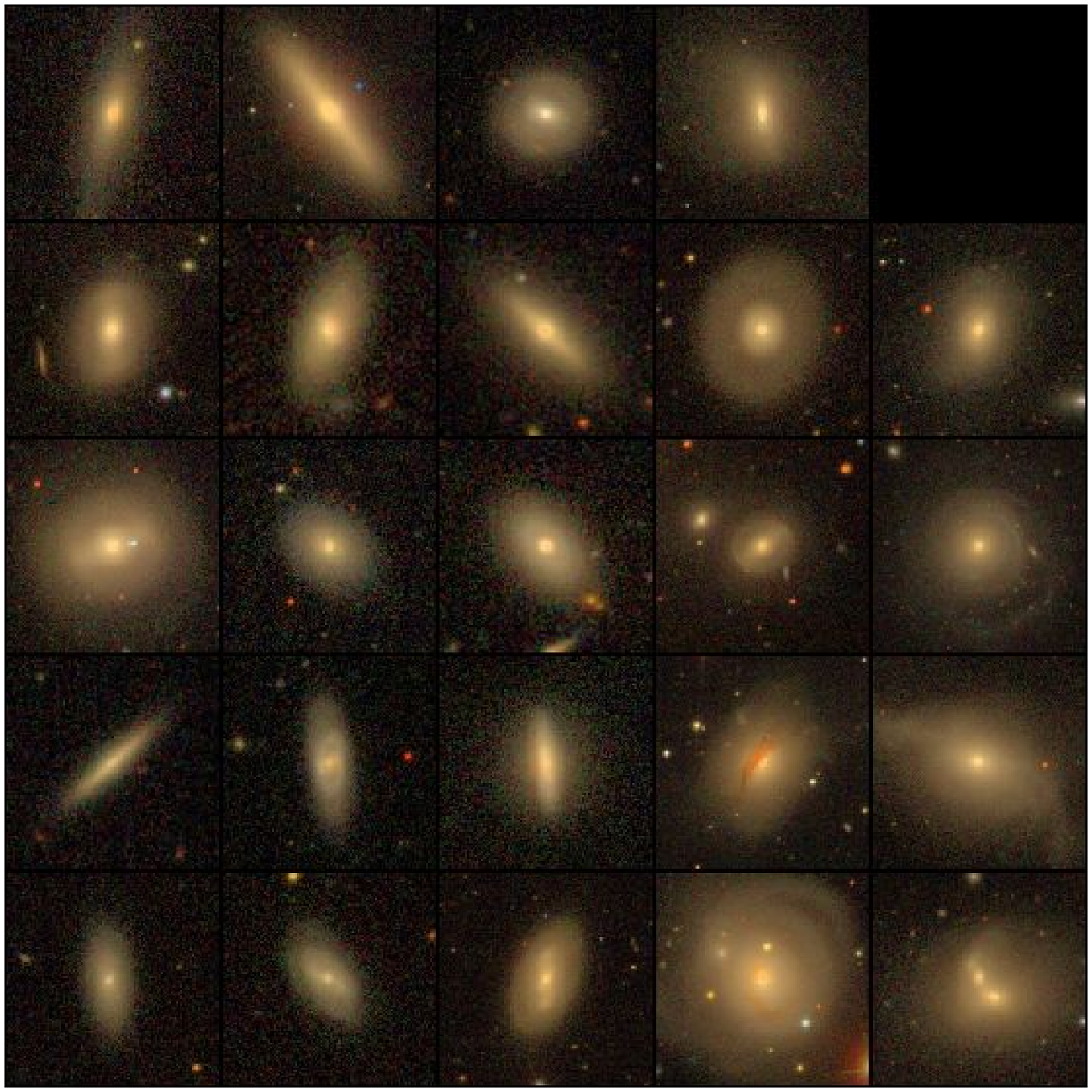}
\caption{\label{lenticulars} SDSS images of lenticular (S0) galaxies, selected 
according to classifications in NED.  The images are sorted by
absolute magnitude in the horizontal direction, ranging between $M_r
-5\log_{10} h\sim -18.5$ and $-22$ from left to right, and
concentration ($r_{90}/r_{50}$) in the vertical direction, ranging
between $2.2$ and $3.8$ from the bottom to the top.  Thus, the
brightest, most concentrated S0s are in the upper right.  The galaxies
shown were selected randomly, but roughly one-quarter were replaced
because their NED classifications were clearly incorrect. We left
cases that could be considered ambiguous in the figure.}
\end{figure}

\begin{figure}
\plotone{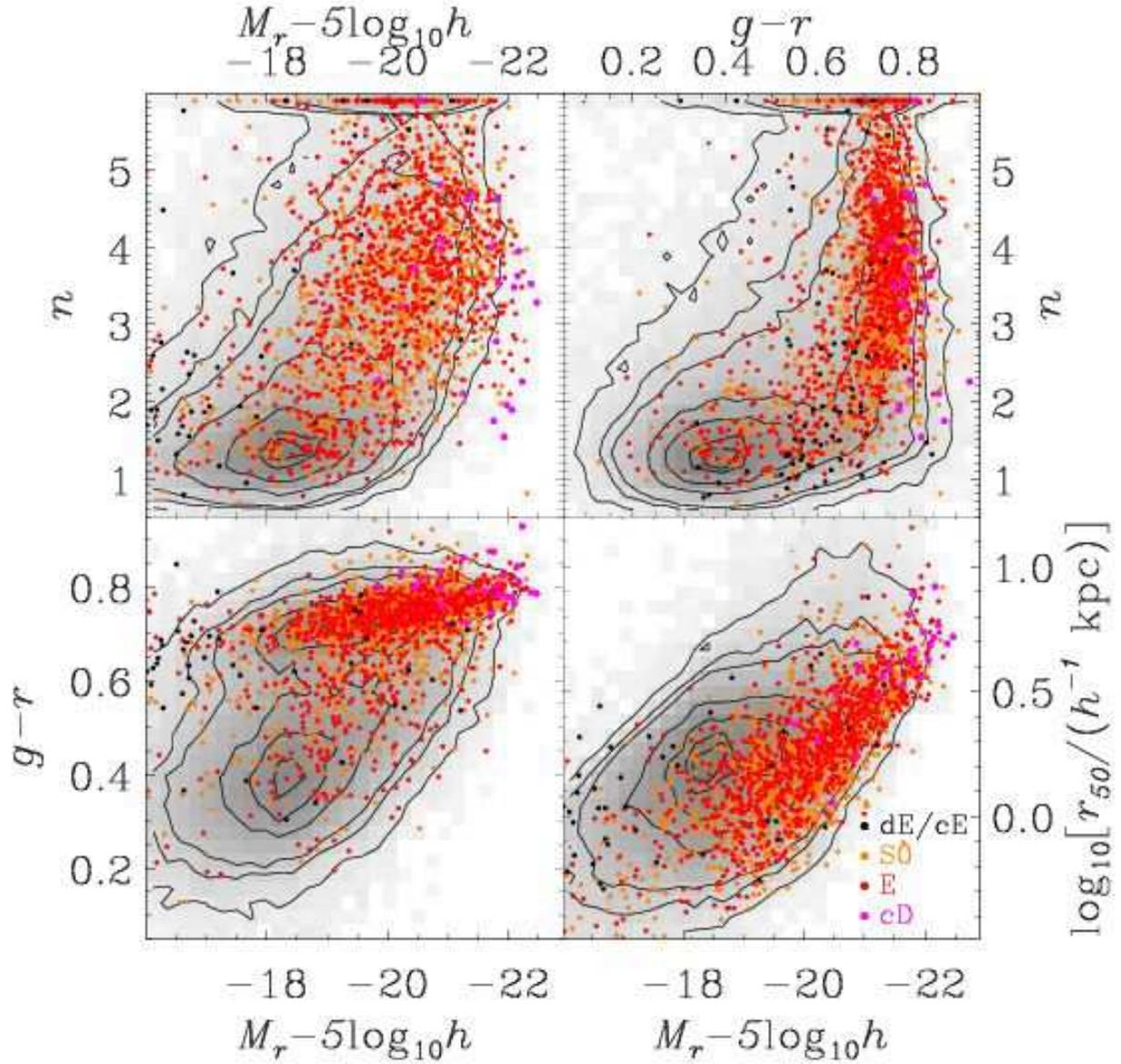}
\caption{\label{manyd-ellipticals} Distribution of 
elliptical and lenticular galaxy types in optical broadband
properties. Similar to Figure
\ref{manyd-spirals}, but now showing S0 galaxies (orange), E galaxies
(red), dE galaxies (black) and cD galaxies (magenta).}
\end{figure}

\begin{figure}
\plotone{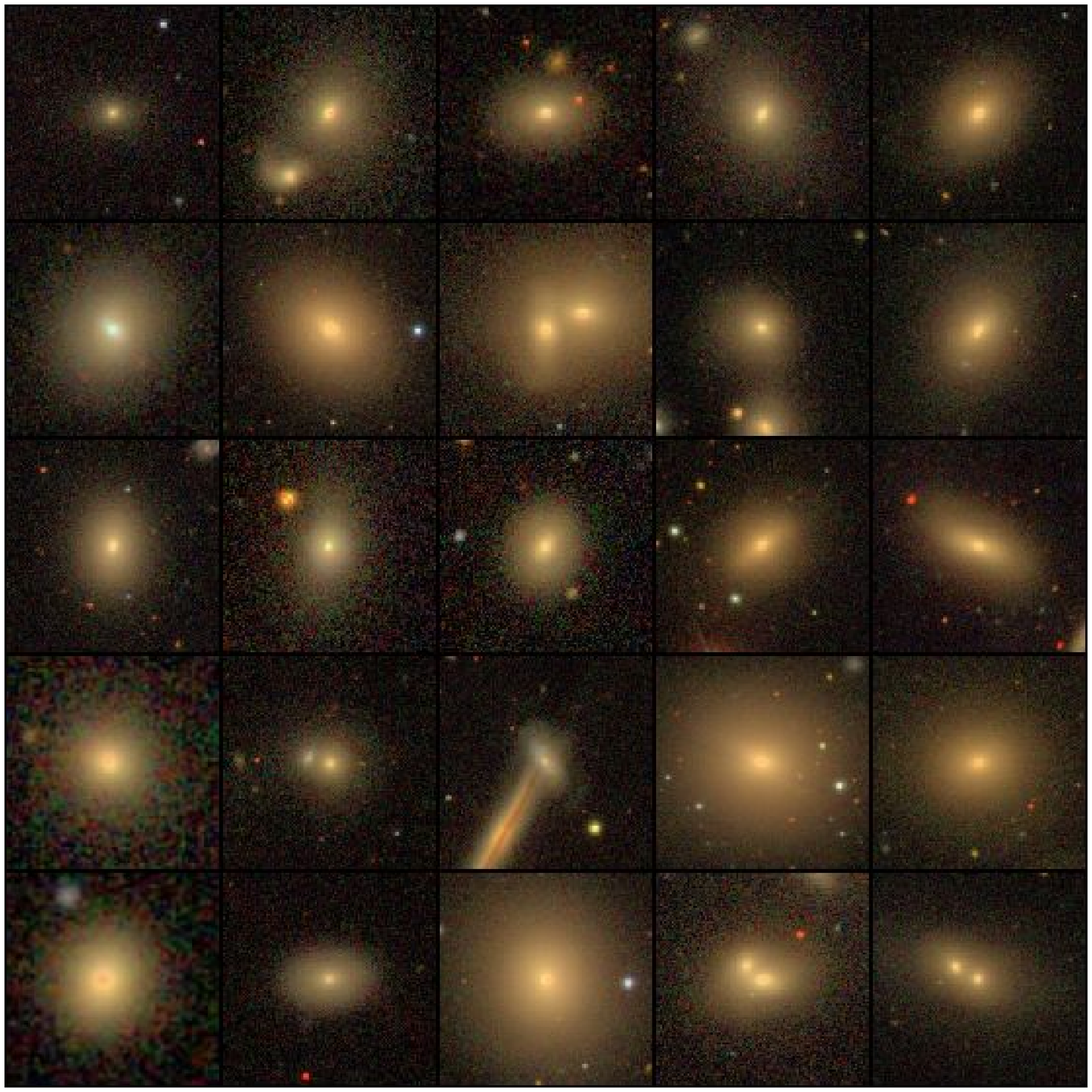}
\caption{\label{ellips} SDSS images of elliptical galaxies, selected 
according to classifications in NED.  The images are sorted by
absolute magnitude in the horizontal direction, ranging between $M_r
-5\log_{10} h\sim -18.5$ and $-22$ from left to right, and
concentration ($r_{90}/r_{50}$) in the vertical direction, ranging
between $2.2$ and $3.8$ from the bottom to the top.  Thus, the
brightest, most concentrated ellipticals are in the upper right.  The
galaxies shown were selected randomly, but roughly one-third were
replaced because their NED classifications were clearly incorrect. We
left cases that could be considered ambiguous in the figure.}
\end{figure}

Although many astrophysicists typically think of ellipticals as de
Vaucouleurs profile galaxies, in fact they have a range of \Sersic\
indices, whose values depend rather strongly on luminosity (as has
been recognized for a long time; \citealt{caon93a, prugniel97a}). The
first row of Figure \ref{props_env} shows this distribution, which
reveals that selecting ellipticals using a strict cut in \Sersic\
index will exclude a significant (and luminosity-dependent) fraction
of viable elliptical galaxy candidates. \citet{Kormendy:2008p1880}
demonstrate this trend most clearly with a careful analysis of
elliptical surface brightness profiles.

These structural trends are mostly independent of environment, as
Figure \ref{props_env} shows (\citealt{blanton05b, park07a}). With the
exception of the most luminous cases, red galaxies of a given
luminosity are no more or less likely to be concentrated when they are
found in dense regions. This result suggests that any process that
turns spiral galaxies into ellipticals acts similarly in dense and
underdense regions.

One shortcoming of the 2D \Sersic\ profile is that when examined
carefully few ellipticals actually have precisely elliptical
isophotes. One way of quantifying this non-ellipticity is to consider
the residuals of the actual isophotes relative to perfect ellipses.
If one Fourier transforms these residuals, the $\cos 4\theta$ term is
of order 1\% of the isophotal radius for many ellipticals
(\citealt{bender87a}). The fractional amplitude of this term $a_4/a$
is defined such that positive values are ``diskier'' than ellipses
while negative values are ``boxier.''  Generally, boxy galaxies are
more luminous than disky galaxies and more likely to exhibit a core
(e.g., \citealt{kormendy89a}; also see \S\ref{fp}).

\subsection{Nuclear properties}
\label{core}

The structure of the central regions of ellipticals are now observable
for large samples using \emph{HST} imaging (\citealt{ferrarese06a,
lauer07b}; see in particular the review of
\citealt{Kormendy:2008p1880}). Although these analyses have varied in
their details, they all find that the central stellar mass surface
density profiles can reasonably be modeled as a power law $\Sigma(r)
\propto r^{-\gamma}$.  As defined by \citet{lauer95a}, galaxies with
$\gamma\sim 0.0$--$0.3$ are classified as ``cores,'' and galaxies with
$\gamma\sim 0.5$--$1.0$ are classified as ``cusps,'' or sometimes
``power-law,'' ``extra-light,'' or just ``coreless.'' There are of
course galaxies in between the extremes (\citealt{rest01a}).

The ellipticals with cores are the most luminous galaxies, with $M_V
-5\log_{10} h<-21$, and correspondingly have distinctly higher global
\Sersic\ indices, boxier isophotes, slower rotation, and more
triaxiality (\citealt{ferrarese06a, lauer07b, Kormendy:2008p1880}).
An oft-used model for the shapes of elliptical galaxy cores is the
``Nuker'' model introduced by \citet{lauer95a}, which is close to a
broken power law. This model does not accurately reflect the surface
brightness profiles of ellipticals at galactic radii larger than
$r_{50}$, which tend to be closer to \Sersic\ profiles.  Consequently,
several researchers (e.g., \citealt{Kormendy:1999p2924, trujillo04a,
ferrarese06a}) have suggested the use of a \Sersic\ model altered to
transition to a power law near the center. These cores are thought to
be scoured by the gravitational effects of binary black holes during
mergers (for a review see \citealt{Kormendy:2008p1880}).

For $-21<M_V -5\log_{10} h<-18$, ellipticals tend to be cuspy
(\citealt{ferrarese06a}).  Relative to the \Sersic\ models, these
coreless galaxies often actually have an ``extra light''
component. The central light typically has isophotes that deviate from
ellipses, being ``diskier'' ($a_4/a > 0$). If cuspy elliptical
galaxies form in mergers, the diskiness of the central light
components suggest that the last of these mergers should have been
``wet,'' with dissipative processes triggering star-formation and
raising the central stellar surface density, and swamping the effects of
black hole scouring (\citealt{Kormendy:2008p1880}).


\subsection{Fundamental plane and dynamics}
\label{fp}

The SDSS survey has provided an unprecedented sample of galaxies with
reliable spectroscopic velocity dispersions, allowing a detailed
analysis of the fundamental plane (FP; \citealt{djorgovski87,
  Faber:1987p2802}) --- the tight relationship among $r_{50}$,
half-light surface brightness $I_{50}$, and velocity dispersion
$\sigma$. \citet{bernardi07a} 
selected nearby early-types according to concentration, color and
emission line strength and found that for a relationship like
$r_{50}\propto \sigma^\alpha I_{50}^\beta$, the best fit is
($\alpha\sim 1.3$, $\beta\sim -0.76$), in only mild disagreement 
with the previous findings of
\citet{jorgensen96} based on more local samples ($\alpha\sim 1.25$,
$\beta\sim -0.82$). 

The FP differs from a na\"ive prediction based on the virial
theorem. From dimensional analysis, $\mathcal{M}=c \sigma^2R/G$, where
$c$ is a structural constant that depends on the orbital structure,
the mass density profile, the surface brightness profile, and an
appropriately weighted mass-to-light ratio, $R$ is a characteristic
radius, and $G$ is the gravitational constant.  Indeed, the rotation
of these parameters into the $\kappa$-space of \citet{bender92a} was
motivated by this virial interpretation.  In this case, for a constant
mass-to-light ratio the FP parameters become ($\alpha=2$, $\beta=-1$),
rather different than the observed ones. Several possible explanations
exist for the deviation from this relation: (a) a variation of the
stellar mass to dynamical mass ratio; (b) a variation of the stellar
mass-to-light ratio; or (c) ``non-homology,'' that is, a variation in
the nature of ellipticals that changes $c$. Most analyses now favor
the first explanation, with a dynamical to stellar mass
ratio that scales as
$(\mathcal{M}_\mathrm{dyn}/\mathcal{M}_\ast)\propto
\sigma^{0.86}$.

First, spectral synthesis modeling suggests that very little of that
variation can be attributed to the stellar mass-to-light ratio
(\citealt{padmanabhan04a, bernardi06a, trujillo04b,
  proctor08a}). Second, detailed modeling of the 2D dynamics using
integral field measurements generally finds a variation of
$\mathcal{M}_{\mathrm{dyn}}/\mathcal{M}_\ast$ that is quantitatively
similar to the virial estimates that assume homology
(\citealt{cappellari06a, thomas07a}).  Third, direct constraints on
mass density profiles with strong gravitational lensing indicate that
the dynamical mass to light ratio is changing (\citealt{bolton08a}).

Nevertheless, as \citet{cappellari06a} notes, it is curious that the
simple virial estimate yields the same mass-to-light ratio trends as
more complex and direct modelling.  After all, the assumptions of
homology must be wrong in detail, as \citet{trujillo04a} point
out. \citet{cappellari06a} further caution that the virial estimate
only works if the de Vaucouleurs estimate of $r_{50}$ is
used --- notwithstanding the fact that the de Vaucouleurs model is
incorrect for many of these galaxies!

The FP is a weak function of environment, as with all other scaling
relationships. For example, \citet{park07a} shows that the
Faber-Jackson relationship does not vary significantly.  The FP
analysis of \citet{bernardi06a} detected a 0.1 mag surface brightness
shift between the highest and lowest density subsamples, which they
estimated was consistent with an age difference of 1 Gyr between
cluster and field ellipticals (see \S\ref{stellar-pops}).

The apparent simplicity of the FP relationship masks a good deal of
variability and complexity in elliptical dynamics. For example,
ellipticals are known to occasionally contain multiple kinematically
distinct components (\citealt{efstathiou82a}), often in the form of
rotating disks around the core (\citealt{bender88a}). In addition, the
kinematics is often misaligned with the photometry or twists as a
function of radius, often interpreted as a sign of triaxiality (e.g.,
\citealt{binney78a, schechter79a}). The recent analysis of SAURON 
data has greatly increased our understanding of the variety and
incidence of kinematic substructures and other features (e.g.,
\citealt{krajnovic08a}).

The FP also masks the importance of rotation to the global dynamics of
many ellipticals.  Traditionally, this importance has been quantified
by the ratio of rotation speed to central velocity dispersion,
$V/\sigma$ (originally due to \citealt{illingworth77a}, but see the
latest discussion of \citealt{binney05a}). Using SAURON data,
\citet{cappellari07a} argue for a different estimator $\lambda_R=\avg{R
|V|}/\avg{R\sqrt{V^2+\sigma^2}}$, where the averages are weighted by
flux.  This measurement requires 2D velocity maps but is apparently a
better tracer of the angular momentum content than $V/\sigma$, which
is more significantly affected by kinematically decoupled cores as
well as projection effects. 

\begin{figure}
\plotone{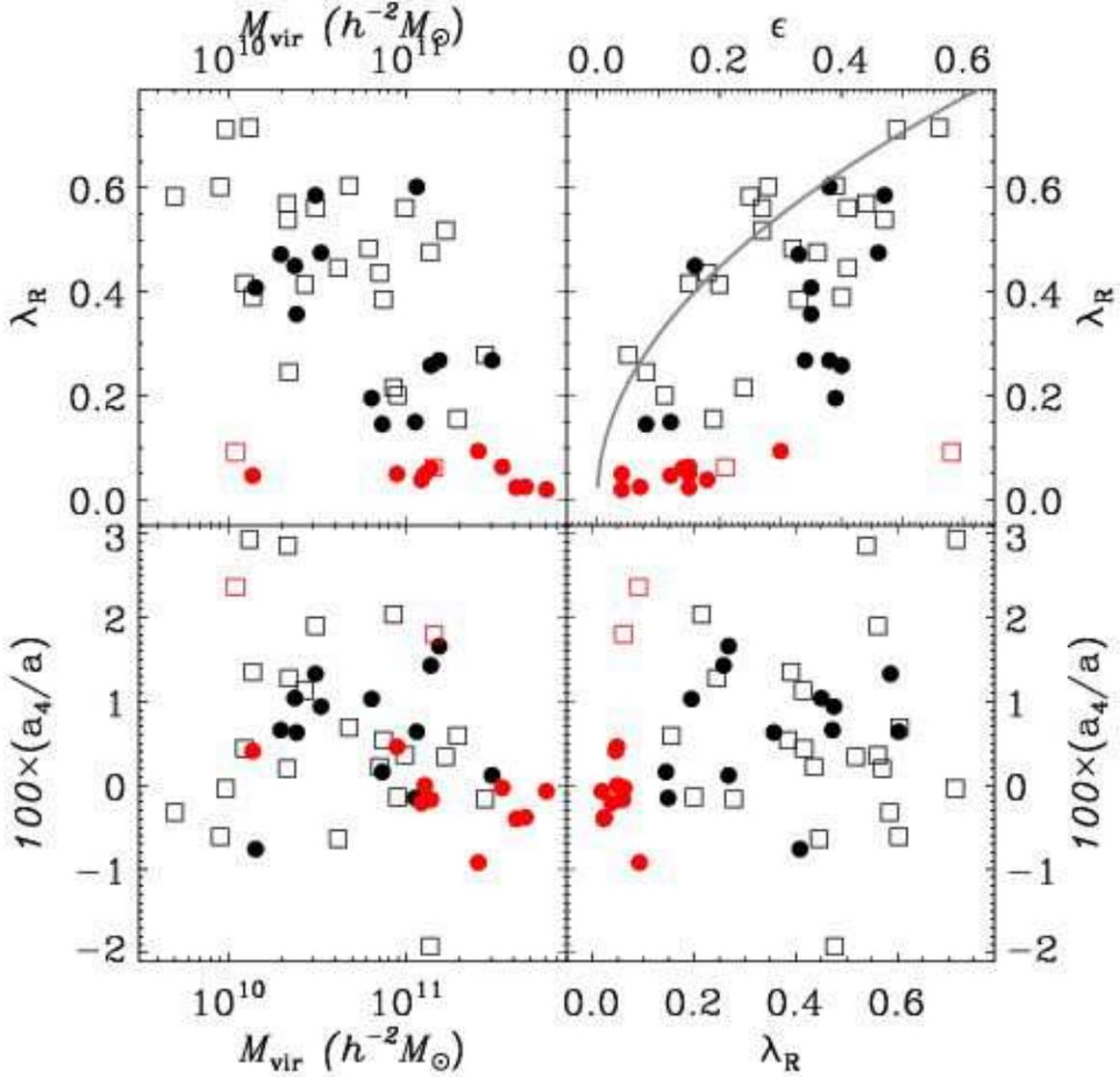}
\caption{\label{emsellem} Distribution of E/S0 galaxy properties from
the SAURON sample of \citet{emsellem07a}. Black symbols indicate
``fast rotators,'' while red points indicate ``slow rotators''.
Filled points indicate optically classified elliptical galaxies, while
open square symbols indicate S0s.  $M_{\mathrm{vir}}$ indicates the
virial mass; $\lambda_R$ indicates the angular momentum content as
defined by \citet{emsellem07a} and described in \S\ref{fp}; $\epsilon$
is the ellipticity within $r_{50}$; and $a_4/a$ is the boxy/disky
parameter (positive is disky, negative is boxy).  The solid grey curve
corresponds to an oblate rotator with an isotropic velocity dispersion
observed edge-on.}
\end{figure}

Figure \ref{emsellem} shows the distribution of $\lambda_R$ from
\citet{emsellem07a}, for the SAURON sample of Es and S0s.  
The upper left panel shows the relationship between angular momentum
content and virial mass $M_{\mathrm{vir}}$.  The red points are the
``slow rotators'' according to the SAURON definition
($\lambda_R<0.1$). For comparison, in the bottom two panels we also
show the distribution of the isophotal shape parameter $a_4/a$
(\S\ref{structure}) as a function of both $M_{\mathrm{vir}}$ and
$\lambda_R$. The upper right panel of Figure \ref{emsellem} shows
$\lambda_R$ versus the observed ellipticity $\epsilon$ within
$r_{50}$.  Overplotted is the approximate curve for an oblate rotator
with an isotropic velocity dispersion, seen edge-on.

The most massive systems have a strong tendency to be the slowest
rotators, the closest to perfect ellipsoids, and the most
axisymmetric.  Relative to the ``fast rotators,'' these slow rotators
tend to have higher masses, flat or falling $\lambda_R$ profiles, less
cuspy centers (\S\ref{structure}), more boxy isophotes, more
kinematically decoupled cores, and greater kinematic misalignments
(indicating triaxiality).  It is difficult to tell with the available
data whether there are some independent relationships among these
properties that are responsible for the others. However, the overall
trend is suggestive of the importance of major dry mergers in the
formation of these systems (e.g., \citealt{hernquist93b}), but perhaps
not minor dry mergers (\citealt{burkert08a}).

The lower mass systems have a stronger tendency to be ``disky'' and
are faster rotators. They tend to have disky isophotes as well as
aligned kinematics and photometry without twists (consistent with the
importance of rotation; \citealt{cappellari07a}). Detailed analysis of
their 2D dynamics suggests that they tend to have anisotropic velocity
dispersions, as their positions in the upper right panel of Figure
\ref{emsellem} suggests (\citealt{cappellari07a}). Interestingly, the
S0s, while always fast rotators, do not appear to be particularly
distinct from fast rotator ellipticals in their dynamics (though the
SAURON analysis is limited to radii $\simless r_{50}$).

Thus, though the FP is simple and appears to remain relatively constant
with environment, there are still only small samples available with
truly detailed 2D dynamics ($\simless 100$ nearby ellipticals total).
Thus, nobody has explored whether these more detailed properties vary
with environment even as the FP remains relatively constant; such a
detection would be an important constraint on formation mechanisms.

In this section (as in this review as a whole) we have focused mainly
on the properties of the luminous galaxies. Just as dwarf disk
galaxies deviate from the Tully-Fisher relation, the dE galaxies
deviate from extrapolations of the FP (\citealt{geha03a, vanzee04b}),
while cE galaxies do not (\citealt{Kormendy:2008p1880}). However, low
luminosity ellipticals do follow a more general, but still regular,
relationship described by
\citet{zaritsky06a}, a ``fundamental manifold'' for spheroids.

\subsection{Brightest cluster galaxies and cD galaxies}
\label{cD}

Among ellipticals, brightest cluster galaxies (BCGs) and cD galaxies
form a special class (\citealt{Morgan:1965p2809, sandage72a}). BCGs
are usually defined as the highest optical luminosity galaxy in any
reasonably massive cluster ($> 10^{14}$ $\mathcal{M}_\odot$).  A large
fraction of BCGs have a larger associated distribution of stars that
extends out into the host cluster, called the ``cD'' envelope for
historical reasons, but sometimes referred to as the ``intracluster
light'' (ICL).

In Figure \ref{manyd-ellipticals}, BCGs and galaxies with cD envelopes
are shown as the magenta dots. BCGs are luminous and massive, with a
log-normal distribution of stellar mass with a mean of
$\mathcal{M}_\ast \sim 2\times 10^{11}$ $h^{-2}$ $\mathcal{M}_\odot$
and dispersion $\sigma_{\ln \mathcal{M}} \sim 0.4$ (\citealt{lin04b,
  hansen07a, yang08b}). 
These luminosities are a function of
the host cluster mass, with $L_K\propto \mathcal{M}_h^{0.2-0.3}$ for
massive clusters (\citealt{lin04a, hansen07a}) 
and a steeper relationship at lower masses (\citealt{yang07a,
  brough08a}).  Thus, as the total halo mass and luminosity rises, the
fractional contribution of the BCG decreases.

\citet{tremaine77a} found hints that BCGs were not merely the
brightest members of a randomly sampled luminosity function for each
cluster.  In particular, the second brightest galaxy is normally at
least 0.8 mag fainter than the first. In contrast, if the luminosities
randomly sampled almost any conceivable distribution, then the
expected magnitude ``gap'' between the first and second ranked cluster
galaxies would be about the same or less than $\sigma_{\ln
  \mathcal{M}}$, or about 0.4 mag (\citealt{vale08a, loh06a}).  This
``gap'' in magnitude suggests an anti-correlation among galaxy
luminosities within the same cluster.

The dry merger, or cannibalism, scenario for the growth of BCGs is
consistent with the last two results. First, the dynamical friction
time scales for galaxies tends to increase as the host cluster mass
increases, explaining why a smaller fraction of the cluster luminosity
is accreted onto the central galaxy as mass increases
(\citealt{cooray05a}).  Second, if the second ranked galaxy is close
in mass to the BCG, it is drawn preferentially to the BCG and will
merge with it. This process will open a gap between the masses and
luminosities of the first and second ranked galaxies (for recent
analyses see \citealt{loh06a, milosavljevic06a}).

Other hints that BCGs are special come from comparing their
fundamental plane
relation to that of other ellipticals.  This 
comparison is complicated by the fact that BCGs are much brighter
than the typical elliptical. \citet{oegerle91a} reported that BCGs
roughly followed the fundamental plane defined by lower luminosity
galaxies.  However, above $M_r -5\log_{10} h\sim -22.3$ and
$\sigma\sim$ 280 km s$^{-1}$, the velocity dispersion of BCGs becomes
a weaker function of luminosity, causing a deviation from the
extrapolation of the 
fundamental plane.  This weakening has been verified in modern
data sets (\citealt{lauer07a, vonderlinden07a, bernardi07a,
desroches07a}). However, at luminosities where BCGs and elliptical
populations overlap, the differences in their velocity dispersions are
$<5\%$ (\citealt{vonderlinden07a, desroches07a}).

Several large studies based on SDSS also indicate that the most
luminous BCGs typically are larger than the radius-luminosity relation
for typical ellipticals would predict ($r_{50} \propto L^\alpha$, with
$\alpha \sim 0.6$).
Either BCGs and other ellipticals lie on different loci in $r_{50}$-$L$
space, or the locus they both live on is curved such that $r_{50}$ is a
stronger function of $L$ at high luminosity. In fact, there is
evidence that both effects exist.  At higher luminosities, the non-BCG
ellipticals appear to deviate from a power-law relation
(\citealt{desroches07a, vonderlinden07a}), with a local power law
$\alpha$ varying from $0.5$ at $M_r -5\log_{10} h\sim-20$ to $0.7$ at
$M_r -5\log_{10} h\sim -24$. Simultaneously, BCGs appear to be larger
than non-BCGs at a given magnitude, but the literature has not
converged on the details. \citet{vonderlinden07a} find that the BCGs
are all 10\% larger on average than non-BCG ellipticals, constant with
luminosity, defining a slightly different fundamental
plane. \citet{desroches07a}, meanwhile, find that the BCGs define a
steeper relationship than the ellipticals, and that at $M_r - 5
\log_{10} h\sim -22$ the typical BCG is smaller than the typical
elliptical. \citet{bernardi07a} also find a steeper slope for BCGs,
but that the relationships converge near $M_r - 5\log_{10} h \sim
-22$.


These results suffer from two major systematic effects First, there
are some significant ambiguities in defining a ``BCG'', particularly
for cluster catalogs (e.g., \citealt{miller05a}) based on an
incomplete redshift survey like the SDSS (even 5\%--10\%
incompleteness matters; see \citealt{vonderlinden07a}). Second, BCGs
are large objects on the sky, and their photometry is not handled
correctly by the SDSS (\S\ref{lf}; \citealt{lauer07a}).  All of the
work cited here either reanalyzes the images (\citealt{bernardi07a,
  desroches07a}) or corrects the catalog parameters in an ad hoc way
(\citealt{vonderlinden07a}).  No cross-comparison of their analyses
has been published.

Many BCGs have an extended distribution of stars that is inconsistent
with a single de Vaucouleurs profile extrapolated from small radii, or
even with a more general \Sersic\ profile, known as a cD envelope or
the ICL. \citet{mihos05a} found that the Virgo Cluster has a
particularly dramatic ICL component, with visible streams and other
features that might suggest a tidal stripping or merging scenario for
its formation.  For a large sample of clusters, \citet{gonzalez05a}
recently showed that detected ICL components are often discernably
separate entities from the host BCGs, with well defined transitions in
the surface brightness profile, axis ratio, and position
angle. \citet{zibetti05a} stacked images of many different SDSS
clusters and statistically detected the ICL component.
\citet{zibetti05a} and a study of individual detections by
\citet{krick07a} both conclude that of order 
5\%--20\% of the total cluster optical light within 500 kpc or so
comes from the ICL; the results of \citet{gonzalez05a} imply even
more. 
If so, the stellar mass in the ICL is comparable
to that in the BCG itself (or possibly 4--5 times the BCG mass if the
analysis of \citealt{gonzalez05a} is correct). Its existence appears
consistent with the merger hypothesis for BCGs themselves.

The dynamics of the ICL component is poorly known in most cases, and
it is unclear to what degree is represents the dynamics of the cluster
as a whole. Some BCGs show the rising velocity dispersion profile at
large radius expected from the cluster dark matter distribution (most
notably those found in \citealt{Dressler:1979p2813, kelson02a}).
However, the large studies of \citet{fisher95a} and \citet{loubser08a}
find that only a minority of cD envelopes have such a profile.
Interestingly, detectable rotation is seen in a number of cD
envelopes, with $V/\sigma \sim 0.3$.

\subsection{Deviations from smooth profiles}
\label{shells}

In addition to boxiness and diskiness, elliptical galaxies often show
other deviations from smooth elliptical isophotes, at least at very
faint levels (\citealt{kormendy89a}). The most common deviation from
smoothness is due to dust features; as we outline in \S\ref{coldgas}, many
elliptical galaxies have a small but detectable amount of cold gas and
associated dust. For example,
\citet{colbert01a} find that about 75\% of ellipticals have detectable
dust extinction in the optical, irrespective of environment.

Elliptical galaxy profiles also often reveal shells or ripples, and
other signs of recent interaction (\citealt{athanassoula85a}).  While
\citet{malin83a} found that about 10\% of all ellipticals had
detectable features, perhaps unsurprisingly it appears that deeper
observations reveal structure in a larger fraction.
\citet{vandokkum05a} find that 70\% of ellipticals have detectable
tidal features down to 27--28 mag arcsec$^{-2}$. For about 20\% of
those detections, a secondary object that appears responsible for the
features is observed. These observations suggest that some growth of
elliptical galaxies occurs through merging; however, precisely how
much is difficult to infer from such observations. \citet{colbert01a}
also searched their sample for tidal features, finding that the
fraction of disturbed ellipticals is a strong function of environment,
more common in isolated regions than in groups and clusters.  Finally,
ellipticals with more fine structure tend to be slightly bluer than
those with less (\citealt{Schweizer:1992p2818}) and perhaps lie on the
bright side of the fundamental plane (\citealt{Michard:2004p2819}).

A final deviation from a purely elliptical profile is that of
isophotal twists. These features, which can be caused by triaxiality
or by tidal effects, appear to be rare: a survey of Virgo ellipticals
using \emph{HST}/ACS reported only seven cases of isophotal twisting (some of
these very marginal or related to central disks) out of 100 observed
early-type galaxies (\citealt{ferrarese06a}).

\subsection{Stellar populations}
\label{stellar-pops}

The spectra of elliptical galaxies are dominated by emission from the
surfaces of stars, typically K giants but comprising some mixture of
stellar types depending on the age, metallicity, and metal abundances of
the stellar population.  For this reason ellipticals all have nearly
the same optical broad-band color, with a weak dependence of color on
galaxy luminosity (and equivalently, stellar mass or velocity
dispersion). This dependence is due to both age and metallicity trends
as a function of mass, as detailed spectroscopic analyses reveal.
\citet{renzini06a} review the subject in detail.  However, in brief,
Balmer absorption lines such as H$\gamma$, H$\delta$ and H$\beta$ tend
to trace age in old stellar populations, while metal-line indices such
as $\avg{Fe}$ and Mg~$b$ yield information about the metallicity and
$\alpha$ abundances
in the stellar atmospheres.

Based on high signal-to-noise optical spectroscopy of morphologically
selected E and S0 galaxies, \citet{thomas05a} show that the
metallicity (or iron abundance) and [$\alpha$/Fe] ratio are both
correlated with velocity dispersion (or mass).  As found previously,
the Mg~$b$ indicator is much more tightly correlated to mass than
$\avg{Fe}$ (e.g. \citealt{worthey98a} and references therein). Similar
results have also been found with large samples of (lower
signal-to-noise) SDSS galaxies (e.g., \citealt{eisenstein03b,
bernardi06a, gallazzi06a, jimenez07a}).

Many ellipticals show evidence for recent star formation in their
optical spectra, and the incidence is higher in lower-mass
galaxies \citep{trager00a, thomas05a}. 
These conclusions have been strengthened by recent observations with
\emph{GALEX}. \citet{yi05a}, \citet{kaviraj07a}, and \citet{schawinski07a}
have all found evidence for recent ($\simless 1$~Gyr) star formation
in $15\%-30\%$ of morphologically selected elliptical galaxies, which
accounts for $1\%-3\%$ of the stellar mass of the galaxy.

\begin{figure}
\plotone{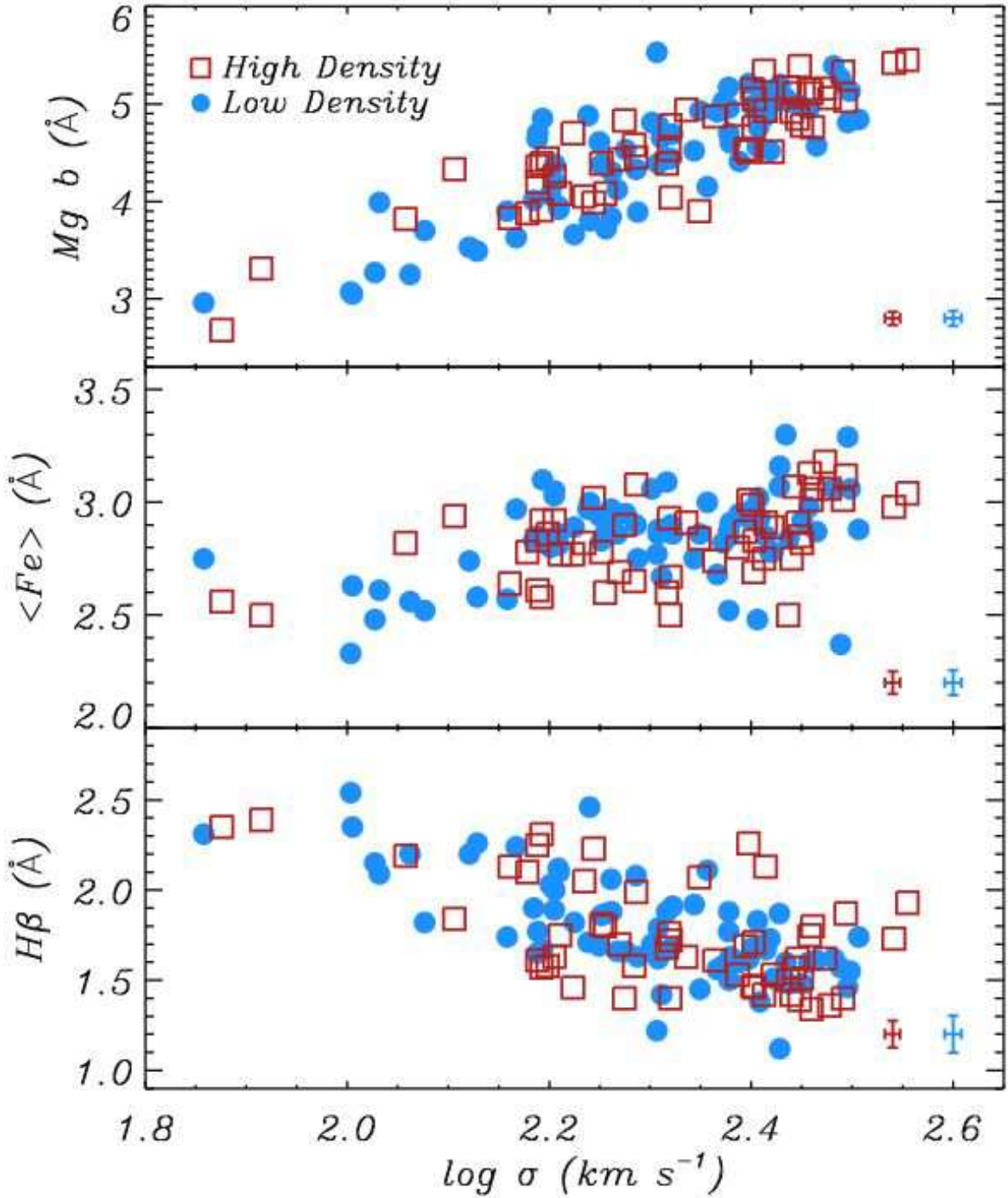}
\caption{\label{thomas} Dependence of E/S0 spectral indices on
$\sigma$ and environment from \citet{thomas05a}. Filled blue circles
are low density points; open red squares are high density points. Top
panel is the Mg $b$ indicator ($\alpha$ abundance), middle panel is
the $\avg{Fe}$ indicator (iron abundance), and bottom panel is
H$\beta$ (an age indicator). Typical uncertainties in the measurements
are shown in the lower right corners.}
\end{figure}

There appears to be some variation of these star-formation histories
with environment. Figure \ref{thomas} shows H$\beta$, $\avg{Fe}$ and
Mg~$b$ as a function of velocity dispersion $\sigma$ for E/S0 galaxies
from \citet{thomas05a}. Filled symbols correspond to galaxies in dense
regions, while unfilled symbols are for galaxies in underdense
regions. They appear to lie on somewhat different loci. According to
\citet{thomas05a}, based on these results the field early-type
galaxies are on average $\sim2$~Gyr younger and slightly more
metal-rich, while both populations show comparable [$\alpha$/Fe]
ratios. \citet{bernardi06a} found qualitatively similar results,
though with a rather different analysis technique.  
\citet{schawinski07a} further found that early-types with recent star
formation were more prevalent in low density environments, even after
controlling for the fact that lower-mass (massive) galaxies are found
preferentially in underdense (dense) regions.  

The nature of elliptical galaxy stellar populations is an old,
storied, and quite controversial one, and the limited space here cannot do it
justice.  However, we do note several investigations that have reached
conclusions in conflict with those described above.  First, recent
\emph{Spitzer} observations of the mid-IR ($9-12$~\micron) spectra of
ellipticals in nearby clusters have found that the vast majority are
consistent with being purely passively evolving systems
\citep{bressan06a, bregman06a}. Second, \citet{trager08a}
find that early-type galaxies in the Coma cluster show the same
mean age as field ellipticals (though less scatter).  The exact nature
of the recent star formation in ellipticals, and its variation with
environment, therefore appears to be in some doubt.



\subsection{Cold gas content}
\label{coldgas}

Elliptical galaxies contain a small amount of cold atomic and
molecular interstellar gas \citep{faber76b}.  When detected, the gas
comprises $\simless 1\%$ of the total mass of the system
\citep{knapp85a}, so it is not a dominant baryonic component.




Early observations showed that the H~{\sc i} detection rates in ellipticals
are several times higher in objects with morphological fine structure
such as visible dust lanes, shells, and ripples \citep{bregman92a,
  vangorkom97a}, which themselves occur more frequently in field
ellipticals \citep{malin83a, schweizer90a}.  This result indicates a
close connection between the gas properties of a galaxy and visible
signs of interaction.

The morphology of the H~{\sc i} gas in early-type galaxies generally
falls into two categories: most have disk- or ring-like structures
with regular kinematics, extending up to $\sim200$~kpc in diameter,
while in others the H~{\sc i} appears in an irregular, tail-like
structure or in individual clouds that are offset from the center of
the galaxy \citep{morganti06a, oosterloo07a}.  \citet{morganti06a}
surveyed a representative subset of the SAURON galaxies and detected
H~{\sc i} emission in $70\%$ of the sample, with gas masses ranging
from $10^{6}$ to $10^{9}~\mathcal{M}_{\odot}$.  They found that
galaxies with H~{\sc i} disks had the most ionized gas emission
\citep[see also][]{serra08a}, but that the H~{\sc i} properties were
uncorrelated with either the age of the stellar population or the
stellar kinematics (fast vs.~slow rotators).
Thus, most early-types seem to accumulate at least some gas,
irrespective of their evolutionary past.  


Surveys of molecular gas in elliptical galaxies have also rapidly
advanced, although the samples remain relatively sparse.
\citet{combes07a} reported that
$28\%$ of the SAURON sample has detectable CO.  Similarly,
\citet{sage07a} detected CO emission in $33\%$ of a volume-limited
sample of field ellipticals; their survey was designed to detect at
least $1\%$ of the gas expected to have been returned to the
interstellar medium by evolved stars in a Hubble time.  
From interferometric mapping, the molecular gas is typically in a
rotationally supported disk $2-12$~kpc in diameter \citep{young02a,
  young05a}.  In many cases these gas disks contain higher specific
angular momentum than the stars, or are counter-rotating with respect
to the stars, suggesting an external origin \citep{young08a}.

The analysis by \citet{combes07a} of CO-rich early-type galaxies also
found that their sample obeyed the Kennicutt-Schmidt relation
(\S\ref{sfr}; \citealt{kennicutt98a}) for disk and starburst galaxies,
but at gas and star formation rate surface densities two orders of
magnitudes lower.  Further evidence for ongoing star formation in
early-type galaxies was presented by \citet{young08b}, who studied a
sample of CO-rich E/S0 galaxies and found good spatial correspondence
between the CO, $24$~\micron, and radio continuum emission, from which
they concluded that the $24$~\micron{} emission is predominantly due
to star formation (rather than AGN or circumstellar in origin).


The nature of the gas in ellipticals is mysterious.  Stellar recycling
arguments suggest that they should contain about ten times the atomic
hydrogen mass that they actually do. Additionally, the gas mass is not
correlated with the stellar mass, which it would be if recycling were
an important source \citep{ciotti91a, sage07a}. Where the recycled gas
goes is an unsolved problem.


%

\section{Interactions, mergers, starbursts, and post-starbursts }
\label{mergersec}

Galaxy mergers are easily visible in about 1\%--2\% of all luminous
galaxies --- frequent enough to suggest that they could be important
to galaxy evolution, but also infrequent enough that real statistical
samples of mergers are only becoming available today. Figure
\ref{mergers} shows a subset of a sample of merging galaxies selected
from SDSS DR6 by eye (Christina Ignarra 2008, private
communication). A wide variety of merger types and tidal features are
evident, including ``dry mergers'' between pairs of red galaxies,
minor mergers, merger-driven starbursts, and large tidal
tails. Although \citet{struck99a} present a detailed classification of
merging systems that attempts to characterize their physical nature,
it has not been applied to any of the newly available large samples.

There are two general ways of counting ``mergers'': searching for
signs in the images, and counting close pairs.  For example, in Figure
\ref{mergers}, we have simply searched for merger signs in the
galaxies by eye. More objective samples have been created by measuring
asymmetry and/or using unsharp-masking techniques
(\citealt{depropris07a, mcintosh08a}).  Though such perturbation-based
samples are pure in the sense of essentially containing only real
physical pairs, they are biased because they require signs of
interaction to be detectable. An alternative technique is to search
for close pairs within about 50 kpc or so.  Close pair samples are
closer to unbiased but even when chosen from redshift surveys include
non-physical pairs (\citealt{barton00a, smith07a}). Although the
distance to the nearest neighbor and the degree of perturbation are
correlated, when
\citet{depropris07a} performed a careful comparison of the two
approaches, they found generally non-overlapping samples. This result
may indicate that close pair samples reveal the pre-merger population
whereas searching for perturbations reveals a later stage.

Using large samples from the SDSS, several groups have studied the
relationship between mergers and star-formation. As the results of
\citet{barton00a} and others had previously indicated
(\citealt{kennicutt87a, kennicutt98b}), close pairs of galaxies show a
factor of $1.5-2$ enhancement in their star-formation rate relative to
a control sample. Equal mass mergers show the clearest signatures
(\citealt{li08a, ellison08a}). \citet{barton07a} use a carefully
calibrated isolation criterion to select pairs that are not part of
larger groups, finding a clearer signature of star-formation
enhancement for such isolated pairs.

Although close neighbors are associated with true starbursts only
rarely, those rare cases still account for about 40\% of the existing
starbursts in the Universe.  Theory suggests that major gas-rich
mergers would create such starbursts --- a consequence of the large
amount of gas that is driven to the center of the merging system
(e.g., \citealt{mihos92a, cox06a}). For these reasons, the most
luminous infrared galaxies, whose luminosity is powered by
dust-obscured star formation, are often associated with major mergers
(\citealt{sanders96a}).

A class of mergers that do not have associated star-formation are the
so-called ``dry mergers'' --- red, gas-poor galaxies merging with
other red, gas-poor galaxies. A classic example of such a merger is
the infall through dynamical friction of an elliptical galaxy in a
cluster onto the central system, building up its stellar mass and
perhaps helping to create a cD envelope (\S\ref{cD}).  Indeed, surveys
at higher redshift indicate that these sorts of mergers play some role
for galaxies on the red sequence (e.g., \citealt{bell05a,
masjedi07a}).  However, at least occasionally mergers of red galaxies
turn out to not be quite ``dry,'' and indeed have substantial gas
content (\citealt{donovan07a}).

A very rare but oft-studied subset of galaxies that may be related to
mergers are the ``post-starbursts.'' Such galaxies can be identified
by their lack of ionized gas producing H$\alpha$ or other emission
lines, indicating no O and B stars, but strong Balmer lines in their
spectra, indicating the presence of A stars. They are often referred
to as ``K+A'' or ``E+A'' galaxies because of their distinct spectral
characteristics (\citealt{dressler83a, zabludoff96a}). Because of the
life-times of A stars, they must have ended star-formation within the
last Gyr or so, and may be undergoing a transformation. It is unknown
whether such transformations result from ram pressure stripping
events, mergers, or something else.  Indeed, post-starburst galaxies
may have more than one formation mechanism.

In Figure \ref{kplusa}, we show one method for selecting such
galaxies, used by \citet{quintero04a}. They fit the full SDSS spectrum
to a sum of an A star template and an old galaxy template, which
results in an arbitrarily normalized ratio of A stars to old stars,
denoted $A/K$. By comparing this ratio to H$\alpha$ equivalent width
they can identify galaxies with no recent star-formation but a
significant contribution of A stars to the integrated
spectrum. Indeed, they observe a spur of such galaxies.

This method of selection differs from previous methods in two ways.
First, they use the H$\alpha$ emission line rather than [O II]
$\lambda$3727 (e.g., \citealt{blake04a}) or a combination of multiple
lines (e.g., \citealt{poggianti04a}).  As \citet{yan06a} show, using
[O II] $\lambda$3727 alone can easily exclude half or more of the K+A
sample, which commonly have weak AGN (\citealt{yang06b}). 
Any method using
H$\alpha$ is preferable in this respect. Second, they use full
spectral fits rather than Balmer line equivalent widths (e.g.,
\citealt{balogh05a}). This
selection alters slightly the ``purity'' of the sample, as any
sufficiently blue continuum will lead to a K+A classification,
regardless of A star content.

These galaxies tend to be high in surface brightness and to be highly
concentrated, yet blue (\citealt{norton01a, quintero04a}), with a
strong possibility that as their stellar population fades they will
become consistent with the red sequence, and
become elliptical galaxies.
\emph{HST} imaging and optical spectroscopy by \citet{yang08a} suggests that
currently they are discrepant from the fundamental plane of
ellipticals --- that is, they are consistent with having the lower
mass-to-light ratios appropriate to their young stellar populations,
and may fade onto the fundamental plane. Most appear
considerably more disturbed than the typical elliptical
\citep{zabludoff96a}, though those disturbances may disappear over
time.

Interestingly, there is very little evidence that these galaxies occur
more frequently in any particular environment, generally following the
environmental trends of late-type galaxies (\citealt{quintero04a,
blake04a, hogg06a, yan08a}). \citet{poggianti04a} claim that in the
Coma cluster there is a population of low luminosity ($M_V > -18.5$)
K+A galaxies associated with dense areas in the hot intracluster
medium. As \citet{yan08a} point out, some K+As may be explained by
interaction with the ICM, but most cannot; the distribution of most
K+As across environment is more consistent with a merger scenario
\citep{zabludoff96a}.

\section{Discussion} 

Over the past decade, astronomers have gathered a huge amount of
information about nearby galaxies.  In some cases these data have
confirmed and made more precise previously known correlations --- such
as the fundamental plane.  In others, they have broken newer ground
--- such as the refinement of infrared indicators of star-formation
and full dynamical modeling of galaxy centers. An overarching theme
has been that the size of the new data sets has allowed us to ask more
sophisticated statistical questions of the galaxy population.  In
particular, our understanding of the effect of galaxy environment on
galaxy properties is now highly refined.

One clear result from these studies is that while a galaxy's
surroundings affect its probability of being a ``late type'' or
``early type'' galaxy, they only secondarily affect the scaling
relations of those types. That is, whereas elliptical galaxies are
more common in groups and clusters than in the field, the processes
that produce them are likely to have been similar no matter where they
are found today. Galaxy formation theorists ought to be trying to
explain this general rule.

Nevertheless, there are several notable exceptions to the rule.  In
particular, the central galaxies in groups and clusters appear to be a
special class (\S\ref{cD}). Very close neighbors ($< 50$ kpc) seem to
affect each other substantially (\S\ref{environs}, \S\ref{mergersec}).
Galaxies in dense regions are at least 0.02 dex more metal rich than
those in the field (\S\ref{yeff}). Elliptical galaxies in dense
regions
may be slightly
older (\S\ref{stellar-pops}, \S\ref{coldgas}), and have a slightly
different 
fundamental plane relationship (\S\ref{fp}). Ellipticals in the field may
also have had a more active recent merger history (\S\ref{shells}).
In addition, while the 
fundamental plane and Tully-Fisher relations, and the relationship
between luminosity and \Sersic\ index, might be constant with
environment, each of those relationships mask more complex structures
--- be they kinematic complexities, bars, or other features. It may
yet be that these more detailed properties depend on environment in
some revealing fashion, even while maintaining the gross
relations. While the theoretical galaxy formation community has yet to
come to agreement on how the broad trends come to be, the observers
ought still to work on teasing out these intriguing deviations.

Another trend that has recently come into focus is the dependence of
mass-to-light ratio on mass. For galaxies around $L_\ast$, there is
generally only a weak relationship; the Tully-Fisher and fundamental
plane relationships are
(approximately) constant mass-to-light ratio.  But clearly the fundamental
plane
relationship shows a slow increase to high masses, which for the BCGs
becomes even more extreme (\S\ref{cD}). Meanwhile, at low masses the
dwarf disk galaxies deviate from the Tully-Fisher relationship, again showing
high mass-to-light ratios; in this case, because they appear to not
have converted much of their cold gas into stars (\S\ref{gas}). Thus,
the halos that seem to produce the most stellar mass per unit dark
matter mass are somewhere around $L_\ast$. Interestingly, such a trend
is exactly what is required for the $\Lambda$CDM mass function, with
its slow cutoff at high mass and steep faint-end slope, to
successfully produce the observed luminosity function
(\S\ref{lfandsf}; \citealt{tasitsiomi04a, seljak05a}). The best
quantification of this trend is that of \citet{zaritsky08a}, who
attempt to construct a dynamical relationship that applies to all
galaxy classes.

What is even more clear, however, is that even with the vast number of
papers written, the available information in the new data sets is
nowhere near being exhausted. Progress is possible because much
(though not all) of the new data is amenable to the sophisticated
techniques that have been used on smaller, less homogeneous surveys.
In particular, a few interesting questions have barely been touched
with the modern data sets and bear further discussion:
\begin{enumerate}
\item The lenticular population has received far too little attention,
either being tossed in along with the elliptical galaxies or ignored
altogether. The E/S0 separation is not an arbitrary one --- the edges,
rings and bars are clear indications of type --- we need well-defined
and consistent methods of identifying S0s in the new massive surveys.
\item The nature and existence of breaks in the stellar profile of
disks, and the incidence of extended UV disks, seem topics ripe for
study in the new massive surveys.  
While there is now a real census of
these galactic features, there has been no study of how they vary with
environment or with other galaxy properties.  
Similarly, there are now reasonably large data sets of bulges,
pseudobulges, and bars, as well as other detailed properties, whose
relationship with close neighbors and the larger scale environment are
largely unknown.
\item The cores of elliptical galaxies are complex and interesting
places. Evaluating the recently proposed role of AGN in these cores to
the shut-off of star-formation in massive galaxies
(e.g. \citealt{croton06a}) may require understanding these cores much
better.  Photometric information alone does not disentangle the
effects of dust and stellar populations, and the stellar population
and dynamical studies available with larger sets of 
IFU observations may yield a less
ambiguous set of models for elliptical cores.
\item More generally, while some crude measures of galaxy 
structural properties exist, there are few efforts to quantify
classical morphological features like spiral arm organization and
pitch angle. Modern techniques in inference applied to modern data
sets should allow such a quantification.  Classical morphology clearly
is important and relevant --- witness its continued popularity among
astronomers studying the detailed properies of galaxies even as
``large data set'' astronomers eschew it. We can and ought to measure
quantitatively the properties it is putatively based upon.
\item New surveys allow us to connect the radio properties of galaxies (such
as neutral and molecular gas content) with their optical properties
(such as stellar mass and age) with large statistical
samples. Although
many interesting findings have resulted from such comparisons, a
comprehensive analysis has not yet been completed.  The stellar
populations are intimately tied to the interstellar gas, without which
our understanding is at best incomplete.  With upcoming blind H~{\sc i}
surveys such as ALFALFA, hopefully this gap will be filled with
respect to atomic gas at least.
\end{enumerate}

\begin{figure}
\plotone{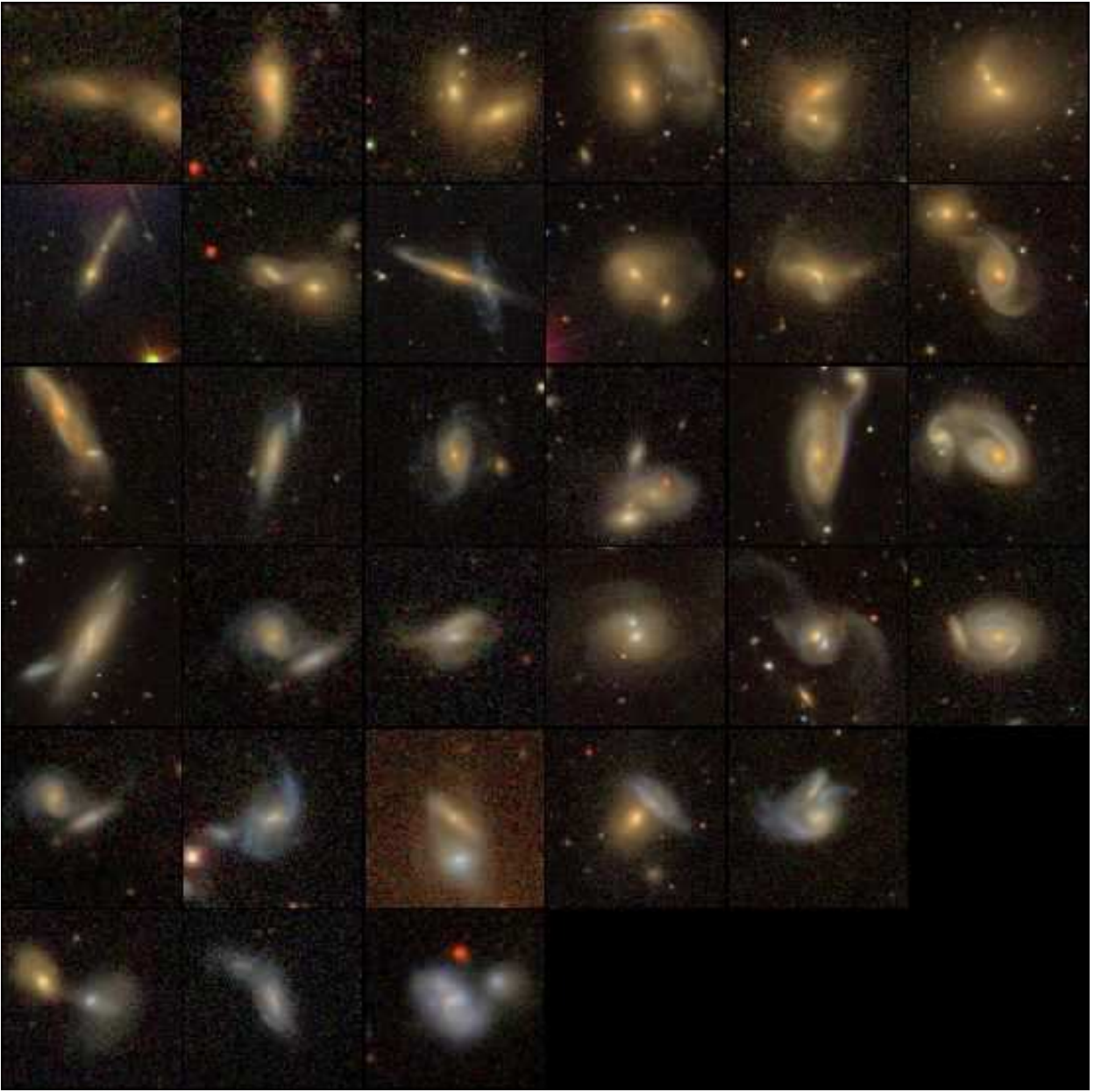}
\caption{\label{mergers} SDSS images of merging galaxies, selected 
by eye from the SDSS DR6 (Christina Ignarra 2008, private
communication).  The images are sorted by absolute magnitude in the
horizontal direction, ranging between $M_r -5\log_{10} h\sim -18.5$
and $-22$ from left to right, and $g-r$ color in the vertical
direction, ranging between $0.2$ and $0.9$~mag from the bottom to the
top.  Thus, the brightest, reddest mergers are in the upper-right.}
\end{figure}

\begin{figure}
\plotone{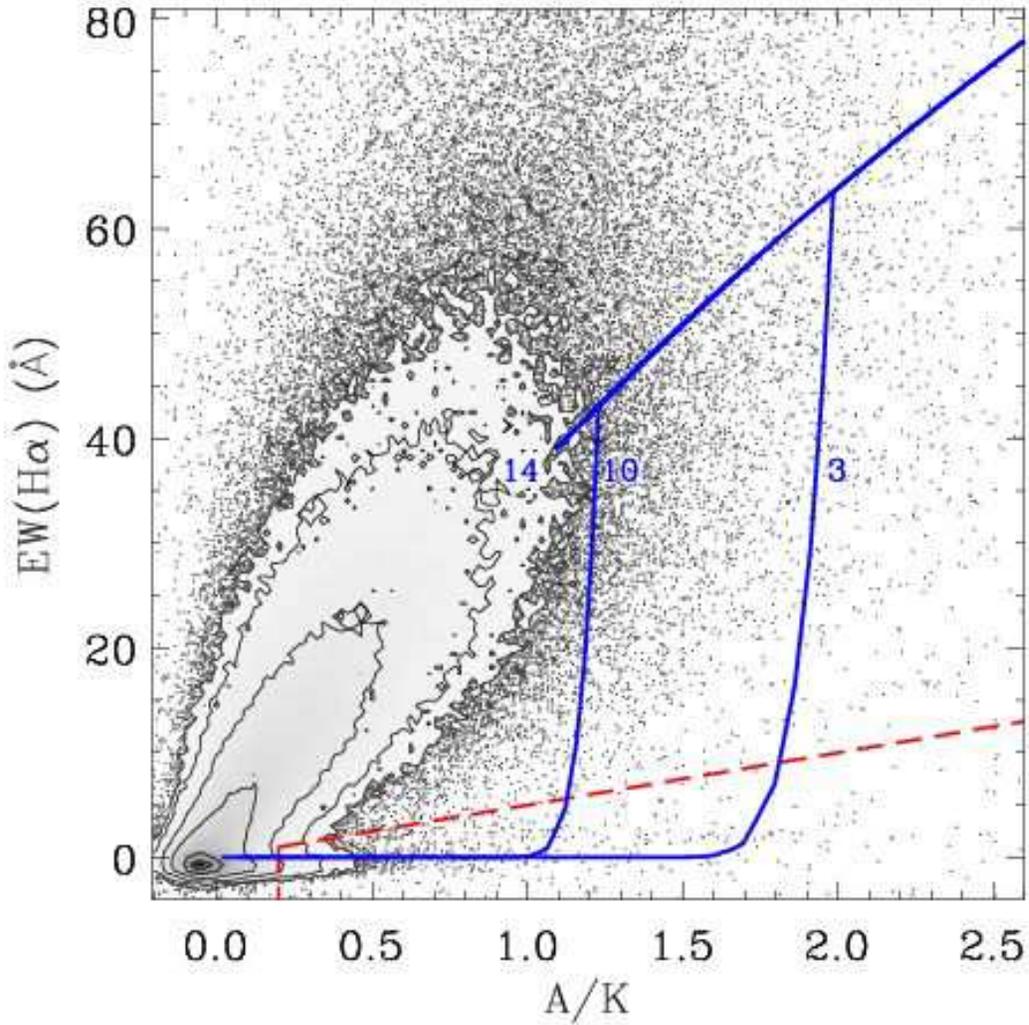}
\caption{\label{kplusa} Distribution of $A/K$ and H$\alpha$ equivalent
width for SDSS galaxies, as determined by \citet{quintero04a}. The
greyscale and contours indicate the density of points in this plane.
Outliers are shown individually. In blue are shown several toy models:
the line labeled ``14'' corresponds to a constant star-formation rate
model over 14 Gyrs; the line labeled ``10'' corresponds to cutting off
that star-formation abruptly at 10 Gyrs; the line labeled ``3''
corresponds to a cutoff at 3 Gyrs. In each of the latter two cases,
the abrubt cutoff results in a post-starburst spectrum.  The dashed
red lines indicate the criteria for selection that \citet{quintero04a}
use.}
\end{figure}

During the course of writing this review, we benefited from
conversations with Adam Bolton, Marla C.~Geha, David W.~Hogg,
Du\v{s}an Kere\v{s}, Michael Hudson, Jill Knapp, and David
Schiminovich.  We would like to thank Christina Ignarra and Alejandro
Quintero for providing some of the data necessary for the Figures
\ref{mergers} and \ref{kplusa}. Finally, we thank our scientific
editor John Kormendy for his extremely useful and challenging
comments.

This research has made use of NASA's Astrophysics Data System and of
the NASA/IPAC Extragalactic Database (NED) which is operated by the
Jet Propulsion Laboratory, California Institute of Technology, under
contract with the National Aeronautics and Space Administration.

Funding for the creation and distribution of the SDSS Archive has been
provided by the Alfred P. Sloan Foundation, the Participating
Institutions, the National Aeronautics and Space Administration, the
National Science Foundation, the U.S. Department of Energy, the
Japanese Monbukagakusho, and the Max Planck Society. The SDSS Web site
is http://www.sdss.org/.

The SDSS is managed by the Astrophysical Research Consortium (ARC) for
the Participating Institutions. The Participating Institutions are The
University of Chicago, Fermilab, the Institute for Advanced Study, the
Japan Participation Group, The Johns Hopkins University, the Korean
Scientist Group, Los Alamos National Laboratory, the
Max-Planck-Institute for Astronomy (MPIA), the Max-Planck-Institute
for Astrophysics (MPA), New Mexico State University, University of
Pittsburgh, University of Portsmouth, Princeton University, the United
States Naval Observatory, and the University of Washington.

The Galaxy Evolution Explorer (GALEX) is a NASA Small Explorer. The
mission was developed in cooperation with the Centre National d'Etudes
Spatiales of France and the Korean Ministry of Science and Technology.

\bibliographystyle{Astronomy}
\bibliography{ccpp.bib}

\newpage
\clearpage

\clearpage

\clearpage

\clearpage

\clearpage

\clearpage

\clearpage

\clearpage

\clearpage

\clearpage

\clearpage

\clearpage

\clearpage

\clearpage

\clearpage

\clearpage

\clearpage

\end{document}